\documentclass[12pt,preprint]{aastex}
\usepackage{amsmath}
\usepackage{graphicx}
\usepackage{enumitem}
\usepackage{color}

\shorttitle{The multi-thermal and multi-stranded nature of coronal rain}
\shortauthors{P. Antolin et al.}

\begin{document}

\title{The multi-thermal and multi-stranded nature of coronal rain}

\author{P. Antolin\altaffilmark{1}, G. Vissers\altaffilmark{2}, T. M. D. Pereira\altaffilmark{2}, L. Rouppe van der Voort\altaffilmark{2}, E. Scullion\altaffilmark{3}}
\affil{\altaffilmark{1}National Astronomical Observatory of Japan, Osawa, Mitaka, Tokyo 181-8588, Japan\\
\altaffilmark{2}Institute of Theoretical Astrophysics, University of Oslo, P.O. Box 1029, Blindern, NO-0315 Oslo, Norway\\
\altaffilmark{3}Trinity College Dublin, College Green, Dublin 2, Ireland}
\email{patrick.antolin@nao.ac.jp}

\begin{abstract}

While the nature of the heating mechanisms in the corona remains elusive their associated cooling is also a poorly known but a far less observationally restrictive subject. In this work, we analyse coordinated observations spanning chromospheric, transition region (TR) and coronal temperatures at very high resolution which reveal essential characteristics of thermally unstable plasmas. Coronal rain is found to be a highly multi-thermal phenomenon with a high degree of co-spatiality in the multi-wavelength emission. EUV darkening and quasi-periodic intensity variations are found to be strongly correlated to coronal rain and especially `showers'. Progressive cooling of coronal rain is observed, leading to a height dependence of the emission. Furthermore, a fast-slow two-step catastrophic cooling progression is found, which may reflect the transition to optically thick plasma states. The intermittent and clumpy appearance of coronal rain at coronal heights becomes more continuous and persistent at chromospheric heights just before impact, in agreement with previous observations above sunspots. This change of character is mainly due to a funnel effect from the observed expansion of the magnetic field at low heights. Strong density inhomogeneities on spatial scales of $0.2\arcsec-0.5\arcsec$ are found, in which TR to chromospheric temperature transition occurs at the lowest detectable scales. The shape of the distribution of coronal rain widths is found to be independent of temperature with peaks close to the resolution limit of each telescope, ranging from $0.2\arcsec$ to $0.8\arcsec$. However we find a sharp increase of clump numbers at the coolest wavelengths and especially at higher resolution, suggesting that the bulk of the rain distribution remains undetected. Rain clumps appear organised in strands. Such structure is not limited to chromospheric temperatures but extends at least to TR temperatures as well, suggesting an important role of thermal instability in the shaping of fundamental loop substructure.  At the smallest detected scales we further find structure reminiscent of the MHD thermal mode. Rain core densities are estimated to vary between $2\times10^{10}~$cm$^{-3}$ and $2.5\times10^{11}$~cm$^{-3}$ leading to significant downward mass fluxes per loop of $1-5 \times10^{9}$~g~s$^{-1}$, suggesting a major role in the chromosphere-corona mass cycle. 

\end{abstract}


\section{Introduction}

The solar corona, the outermost layer of the solar atmosphere, has an average temperature above a million degrees, a fact that has puzzled astrophysicists for decades. The processes responsible for this heating are still unknown, mostly due to the very short time and spatial scales on which they operate. However, not everything in the corona is hot. Structures at chromospheric temperatures are clearly present in the solar corona. Evidence for this has been found for more than a century, since the first detection of prominences in the solar atmosphere \citep{Secchi_1875leso.book.....S}. In the last 40 years, especially in the last few years with the advent of high resolution instruments, smaller chromospheric structures with a more pervasive character have been shown to occur intermittently in the corona above active regions \citep{Kawaguchi_1970PASJ...22..405K,Leroy_1972SoPh...25..413L,Schrijver_2001SoPh..198..325S,Antolin_Rouppe_2012ApJ...745..152A,Antolin_etal_2012SoPh..280..457A,Ahn_2014SoPh..289.4117A}. This material is known as coronal rain, partially ionised, dense clumpy structures catastrophically cooling in the corona and accreting towards the solar surface. Its occurrence is considered as the direct observational consequence of the thermal instability mechanism, in which radiative losses locally overcome the heating processes. 

The presence of prominences in the corona and especially the recent detection of coronal rain with a more pervasive character entails fundamental aspects of the interaction of plasmas and fields in low $\beta$ environment in which thermal instability plays a central role. This mechanism is intrinsically linked to the characteristic cooling aspects of the heating mechanisms and provides far less strict detection possibilities due to the longer timescales and an enormous gain in spatial resolution when observing in wavelengths associated with cool emission. The study of coronal heating through associated cooling therefore provides unique advantages. 

A big unknown in the physics of the solar corona is what percentage of coronal plasma is thermally unstable. This is also of significant importance for stellar coronae, since high speed accretion of chromospheric plasma can produce characteristic profiles in their UV and optical range spectra, both from the falling plasma and from the impact into the lower atmosphere \citep{Ayres_2010ApJ...723L..38A,Reale_2013Sci...341..251R,Reale_2014ApJ...797L...5R}. Based on numerical simulations, such conditions are expected if the heating is mainly concentrated towards the footpoints \citep{Antiochos_1999ApJ...512..985A,Karpen_etal_2001ApJ...553L..85K,Luna_etal_2012ApJ...746...30L}, a condition often reported in active regions, and are to a lesser extent related to the temporal character of the heating, i.e. whether it is impulsive or steady \citep{Muller_2003AA...411..605M,Muller_2004AA...424..289M}. Forward modelling of numerical results and comparison with EUV light curves obtained in observations have helped answering this question, but a debate exists on whether the heating requirements setting loops in a thermal non-equilibrium state fully match the general observed characteristics of active region loops \citep{Mok_etal_2008ApJ...679L.161M,Klimchuk_2010ApJ...714.1239K,Peter_etal_2012AA...537A.152P,Lionello_2013ApJ...773..134L,Klimchuk_2014arXiv1410.5660K}. Coronal rain observations can help answer this question by estimating the coronal volume and mass flux involved in these events. A first attempt at this was performed by \citet{Schrijver_2001SoPh..198..325S} using channels from the \textit{Transition and Coronal Explorer} (\textit{TRACE}), which lead to an occurrence interval of several days for one active region loop to undergo catastrophic cooling. Based on H$\alpha$ data from the CRISP instrument at the \textit{Swedish 1-m Solar Telescope} (\textit{SST}), \citet[][hereafter Paper~1]{Antolin_Rouppe_2012ApJ...745..152A} reduced the occurrence frequency to half a day or less, and estimated the coronal rain mass flux to be about one third of that uploaded into the corona from spicules, thus stressing an important role in the chromosphere-corona mass cycle.  

Paper 1 predicted that most of the rain material may pass undetected in coarse resolution instruments due to average clump\footnote{also referred to as `blobs'} widths of 300~km (and lengths of 700~km or so). Condensations were reported to be clustered in space into groups of correlated events, which indicates that the properties of coronal heating can be correlated over a significant spatial scale and that individual strands (the theoretical internal components of coronal loops) are not necessarily uncorrelated in their evolution. However, while most condensations were found to be correlated in time, only few events, termed `showers' in Paper~1, were found to occur close enough, which would produce large enough EUV absorption detectable with the coarser resolution of coronal instruments. Still, the limit cycles that thermally unstable loops experience, predicted from numerical simulations, should be visible observationally as periodic variations of the EUV intensity (which would then be correlated to the appearance of coronal rain). Such behaviour has recently been reported and seems to fit well into the thermal non-equilibrium scenario \citep{Froment_2014cosp...40E.903F} . 

Since it is chromospheric (cool) plasma occurring in the corona, coronal rain constitutes the highest resolution window into the substructure of loops, and provides a tool for revealing the global magnetic field topology. First evidence from this was provided in Paper~1, where tracking of rain over a decaying active region allowed us to calculate the fall angle of the clumps, and therefore to reveal the large scale morphology of the magnetic field and distinguish several loop families by tracing the clumps along the legs. The coronal rain condensations were reported to be organised locally as strand-like structure with average widths around 300~km and lengths around ~700 km, with nonetheless broad distributions down to 100 km and 200 km, and up to 700 km and 2400~km respectively for widths and lengths. The multi-strand substructure of loops was also reported in on-disc observations of coronal rain with CRISP in \citet{Antolin_etal_2012SoPh..280..457A} and more recently in \citet{Scullion_2014ApJ...797...36S}. The latter work extends such findings to the coronal rain produced in post-flare loops and to the apexes of loops. In that work the distribution of widths peaks at 100~km, clearly indicating that the bulk of the distribution is unresolved due to the lack of resolution. Interestingly, the lengths of some of the observed condensations were reported to span up to 26000~km, further underlining the global coronal magnetic field tracing possibilities offered by coronal rain. 

The strand-like structure revealed by coronal rain along loops appears similar to the thread-like structure observed in prominences of certain types \citep[e.g., active region prominences,][]{Okamoto_2007Sci...318.1577O}. A relevant question is therefore whether such structure is strongly dependent on thermal instability, or whether such structure also exists at coronal temperatures under equilibrium conditions, in accordance with the debated multi-strand scenario in loops \citep{Klimchuk_2006SoPh..234...41K,Brooks_2012ApJ...755L..33B,Brooks_2013ApJ...772L..19B,Peter_2013AA...556A.104P}. The loss of pressure produced by thermal instability produces local accumulation of material leading to condensations. Given a thermally unstable loop, the continuous occurrence in the same spot of thermal instability, combined with the produced strong flows (or other possible siphon flows) in a low $\beta$ environment such as the corona can easily lead to strand-like structure, as shown in multidimensional MHD simulations \citep{Fang_2013ApJ...771L..29F}. In this paper we show that even in this scenario it is reasonable to expect that such strand-like structure is not locally confined to the thermally unstable regions (and thus only confined to the low temperature plasma in the loop), but extends along the entire loop and therefore to the hot coronal plasma as well. Therefore, even if a loop does not tend to be originally multi-stranded, it is possible that such organisation is attained due to thermal instability and maintained thereafter. Such scenario, for which we provide support in this paper \citep[also theoretically supported by ][]{Low_2012ApJ...757...21L}, has not been considered before and may significantly impact on the evolution of the coronal loop.

It can therefore be expected that a multi-thermal structure accompanies a multi-strand organisation of the plasma in thermally unstable loops. This is shown in \citet{Scullion_2014ApJ...797...36S}, where loops observed in the coronal filters of AIA are co-spatial with the H$\alpha$ condensations. The lack of resolution of AIA unfortunately does not allow us to see the details and spatial scales at which such multi-thermality can exist. In this work, with the help of the high resolution of the Interface Region Imaging Spectrograph (\textit{IRIS}), we extend this result.

We present two datasets that combine multi-wavelength instruments spanning chromospheric, transition region and coronal temperatures. \textit{SST}/CRISP , \textit{Hinode}/SOT, \textit{IRIS}/SJI and \textit{SDO}/AIA are used to reveal the multi-strand and multi-thermal aspects of coronal rain at unprecedented detail. The paper is organised as follows. In Section~\ref{obs} the observations are presented. In Sections \ref{temp} and \ref{morph} the temperature and morphology of the observed condensations are presented, respectively. Results are discussed in Section~\ref{discuss} and conclusions are given in Section~\ref{end}.

\section{Observations}\label{obs}

\subsection{Data processing}

\begin{figure}[h]
\epsscale{1.}
\plotone{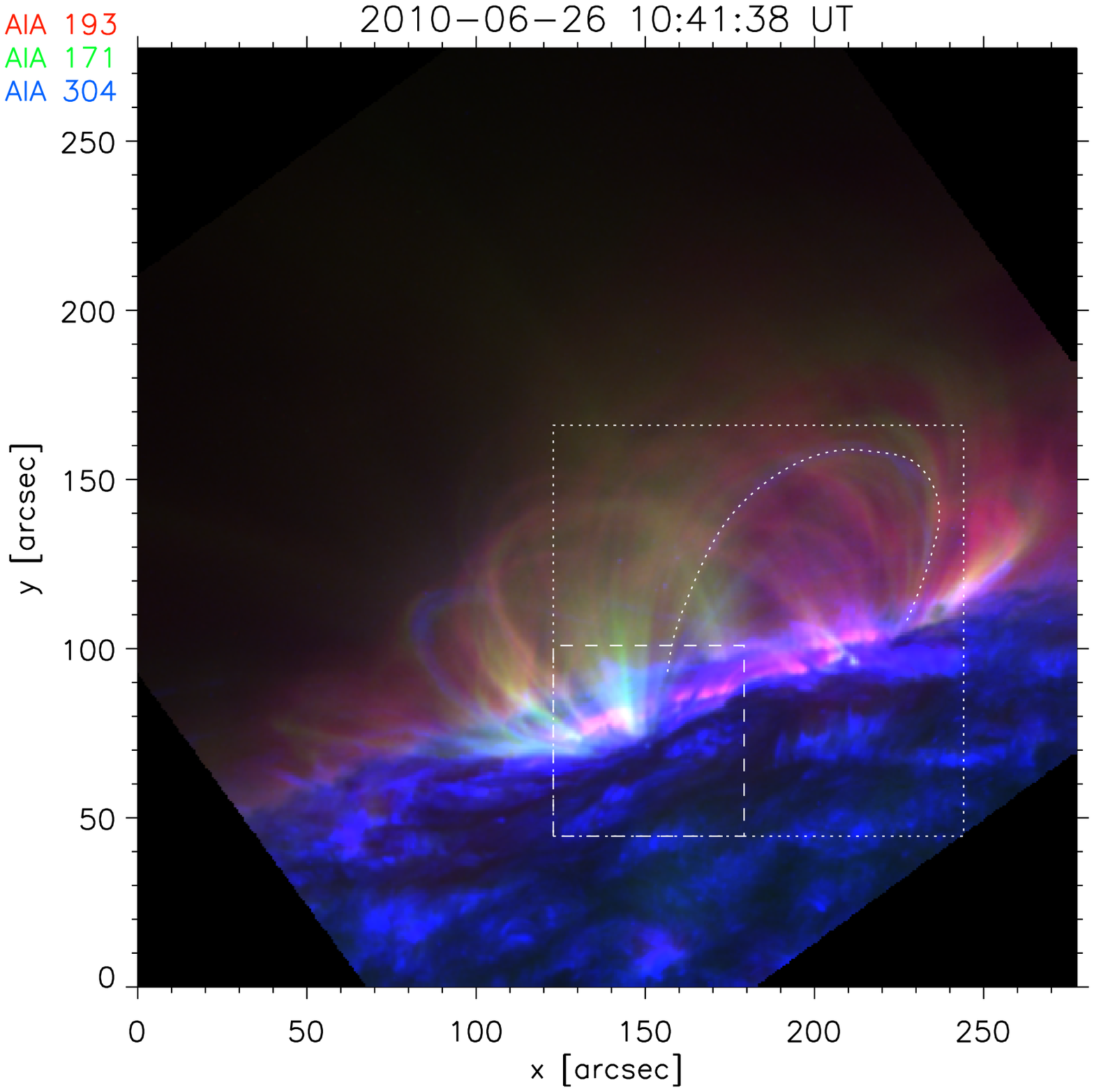}
\caption{Composite image from \textit{SDO}/AIA combining three AIA filters, 193 (\textit{red}), 171 (green) and 304 (blue), showing dataset~1: AR 11903 at the East limb on the 26th of June 2010. The dotted square denotes the field-of-view of Fig.~\ref{fig3}. The dotted loop corresponds to loop~2 studied in Section~\ref{euv2ha}. The dashed square denotes the field-of-view of the \textit{SST}. An animation of this figure is available in the online material as Movie~1. 
\label{fig1}}
\end{figure}

\begin{figure}
\epsscale{1.}
\plotone{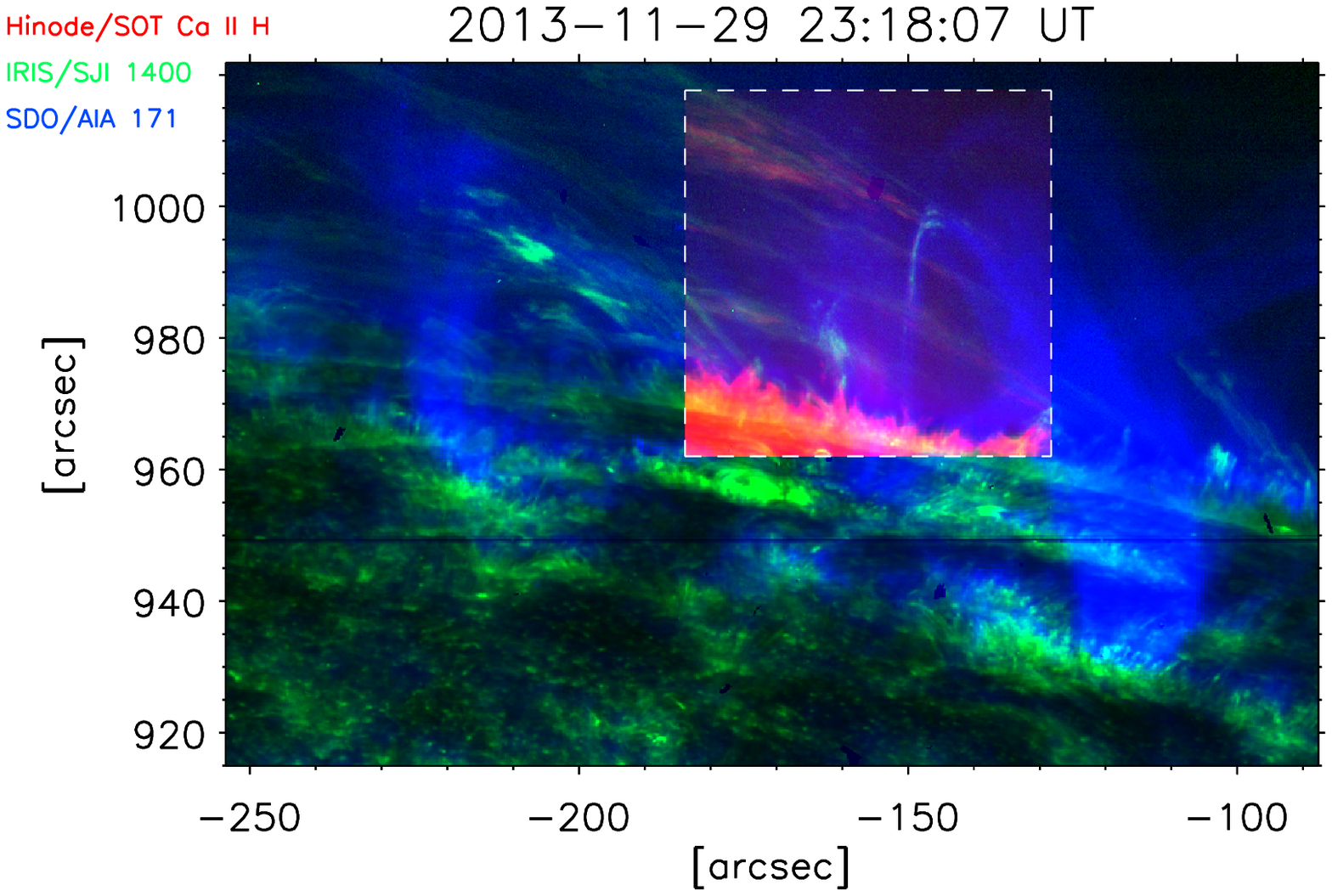}
\caption{Composite image combining AIA 171 (blue), \textit{IRIS}/SJI 1400 (green) and \textit{Hinode}/SOT \ion{Ca}{2}~H (red), showing dataset~2: AR 11903 at the West limb on the 29th November 2013. The dashed square denotes the field-of-view covered by \textit{Hinode}/SOT. The loop with coronal rain towards the centre right of the square corresponds to the loop studied in this dataset. An animation of this figure is available in the online material as Movie~2. 
\label{fig2}}
\end{figure}

In this work we analyse two different datasets corresponding to solar limb observations above active regions. The first dataset combines data from the Swedish 1-m Solar Telescope \citep[\textit{SST};][]{Scharmer_2003SPIE.4853..341S} with the CRISP spectropolarimeter and the AIA instrument \citep{Lemen_etal_2011SoPh..tmp..172L} of the \textit{Solar Dynamics Observatory} (\textit{SDO}), dates from 26 June 2010 from 10:03 UT to 11:40 UT, and was centred on AR 11084 at $[x,y]=[-875,-319]$. The second dataset combines the SOT telescope on board of \textit{Hinode} \citep{Tsuneta_2008SoPh..249..167T}, the Slit-Jaw Imaging (SJI) instrument of \textit{IRIS} \citep{DePontieu_2014SoPh..289.2733D} and \textit{SDO}/AIA, dates from 29 November 2013 from 22:30 UT to 23:30 UT and focused on AR 11903. \textit{IRIS}/SJI was centred at $[x,y]=[944,-264]$ fully containing the \textit{Hinode}/SOT FOV, centred at $[x,y]=[959,-220]$. Context images for these coordinated observations are presented in Figs.~\ref{fig1} and \ref{fig2}, where different colours denote different wavelengths and the different field-of-views (FOV) of each instrument can be compared. 

We mainly focused on the \textit{SDO}/AIA filters 304, 171 and 193, corresponding to a range of transition region to coronal temperatures usually spanned by thermally unstable plasmas. The upper photospheric filters 1600 and 1700 were also used for co-alignment purposes (see Section~\ref{coali}). Level~1.5 data from the \textit{SDO}/AIA was used, obtained through the normal calibration routines in SolarSoft. Full disc images are provided from AIA with a cadence of 12~s, an image scale pixel of 0$.\!\!^{\prime\prime}$6 pixel$^{-1}$ and a spatial resolution of 1$.\!\!^{\prime\prime}$3 - 1$.\!\!^{\prime\prime}$7. 

\textit{IRIS} combines a Slit-Jaw imager (SJI) and a spectrograph (SG). The \textit{IRIS} data used here comprises only the SJI instrument since the \textit{IRIS} slit was not located on the region of interest. Therefore, in this work only the coronal rain imaging capabilities of \textit{IRIS} are shown. The dataset corresponds to level 2 data from a sit-and-stare observation combining the SJI 1400, 1330 and 2796 filters, dominated respectively by the \ion{Si}{4} transition region lines 1393.78~\AA\ and 1402.77~\AA\ (formed around $\log T=4.8$), the  \ion{C}{2} chromospheric lines 1334.53~\AA\ and 1335.71~\AA\ (formed around $\log T=4.3$), and the \ion{Mg}{2}~k chromospheric line 2796.35~\AA\ (formed around $\log T=4$). \textit{IRIS}/SJI provides a FOV of $175\arcsec\times175\arcsec$, an image scale pixel of 0$.\!\!^{\prime\prime}$166 pixel$^{-1}$ and a spatial resolution of 0$.\!\!^{\prime\prime}$33 (for the FUV) - 0$.\!\!^{\prime\prime}$4 (for the NUV). The cadence for this observation is 36.5~s in each filter. 

The CRisp Imaging SpectroPolarimeter \citep[CRISP;][]{Scharmer_2008ApJ...689L..69S} at the \textit{SST} sampled the H$\alpha$ spectral line at 41 line positions with 0.085~\AA\ steps ($3.9~$km s$^{-1}$), in the spectral window [$-1.7, 1.7$]~\AA\ ($\simeq\pm78$~km~s$^{-1}$) with a cadence of 9~s. The FOV of the \textit{SST} for the present observation is $53\arcsec\times57\arcsec$. The images benefited from the \textit{SST} adaptive optics system \citep{Scharmer_etal_2003SPIE.4853..370S} and the image restoration technique Multi-Object Multi-Frame Blind Deconvolution \citep[MOMFBD;][]{VanNoort_etal_2005SoPh..228..191V}. Although the observations suffered from seeing effects, most of the images are close to the theoretical diffraction limit for the \textit{SST} at the wavelength of H$\alpha$:  $\lambda/D\simeq0.\!\!^{\prime\prime}$14. We employed the same reduction procedure as in Paper~1 which used early versions of parts of the data pipeline CRISPRED \citep{DeLaCruz_2015AA...573A..40D}. 

\textit{Hinode}/SOT recorded filtergrams in the \ion{Ca}{2}~H band (formed around $\log T=4$) at a cadence of 4.8~s. The image scale pixel is 0$.\!\!^{\prime\prime}$109 pixel$^{-1}$ and the spatial resolution is $\lambda/D\simeq0.\!\!^{\prime\prime}$2. The SOT FOV for this observation is $55.8\arcsec\times55.8\arcsec$. Processing of data was carried out through the normal calibration routines in SolarSoft. 

\subsection{Co-alignment procedures}\label{coali}

The co-alignment of \textit{SST}/CRISP and \textit{SDO}/AIA data for the 26 June 2010 first consisted in determining the solar-$x/y$ coordinates and locate the solar north direction for each CRISP frame. We then eliminated the \textit{SST} drift by fixing each frame on the sunspot of AR 11084, at the centre of the \textit{SST} FOV, to reduce the frame-to-frame motion of the loop's footpoints (and therefore of the loops themselves). The location of the solar limb for each frame was also determined, which allowed to apply a gradient filter reducing the intensity of on-disc features for better visualisation of the off-limb structures. \textit{SDO} data was then cropped and interpolated to the exact \textit{SST} FOV. Small misalignments still remain, which are then corrected through cross-correlation using photospheric bright points (with long enough lifetime) and the solar limb location (using the AIA filters 1600 and 1700, in combination with the far wing positions of H$\alpha$, which mostly show the photosphere).

For the SOT-SJI-AIA co-alignment, we first removed the slow drift of SOT by co-alignment of the images through tracking the movement of the limb. The cumulative offsets from this drift were applied to the time series, resulting in a rigid co-alignment on the order of one pixel or less. The \textit{IRIS}/SJI images do not have this slow drift, but occasional wobble shifted the images a few pixels (this is normal during eclipse season). Such shifts were measured by applying cross-correlation techniques and applied to the time series. Then, to co-align the images from different instruments the following procedure was used. First, the AIA images were co-aligned and cropped to the SJI field of view by identifying common bright points in the AIA 1600 filter and the SJI 1400 filter. Afterwards, both AIA and SJI images were interpolated to SOT's image pixel scale and were co-aligned using common features on the SJI 1400 filter and SOT's \ion{Ca}{2}~H filter.

\subsection{Methods}\label{methods}

Due to the complex interplay of forces in loops coronal rain dynamics can be extremely complex, especially when observing at sub-arcsecond resolution. Furthermore, their thermodynamic state is also expected to change, as we show in this paper, and as has been also indicated in \citet{Harra_2014ApJ...792...93H}. A consequence of this is that it is very difficult to follow a single rain clump (at high resolution) over long distances. The clumps will usually split, merge, elongate along their paths. For these reasons, especially for the CRISP data where myriads of small clumps are observed and for which the line-of-sight (LOS) superposition is significant, individual tracking is very difficult and time consuming. For automatically detecting rain in the CRISP data we have therefore opted for perpendicular cuts across the loops at positions where rain is observed, rather than longitudinal paths along individual clumps. Detection is done through semi-automated procedures verifying specific conditions for the clumps. These consist especially of contrast thresholds (a rain pixel needs to be significantly brighter or darker with respect to the background, depending on whether the detection is realised off-limb or on-disc, respectively; all frames with bad seeing are then removed), intensity variation thresholds (a rain pixel must present an average change in intensity above a specific threshold within a reasonably small time interval), and size restrictions (a clump must span at least 3 rain pixels in the direction perpendicular to the direction of propagation). Once the rain pixels are identified the clump sizes are calculated with Gaussian fits along and across the direction of propagation. Therefore, one size value for a clump is the average over several tens or hundreds of measurements. The results are also tested for specific conditions, such as size thresholds, small fit errors, consistent widths (a clump's width must not differ significantly over a short distance along the axis of propagation) and so on. For rain detection with CRISP we have further opted for using Doppler intensities rather than the clump intensities alone, since detection is improved in this way. The Doppler intensity for a rain pixel at position $x,y$ and time $t$ is defined as $I_{\mbox{Dopp}}(x,y,\lambda,t)=I(x,y,\lambda,t)-I(x,y,\lambda_0-(\lambda-\lambda_0),t)$, where $\lambda_0$ is the theoretical wavelength value at rest.

For the rain observed with SOT, SJI and AIA we have opted for the `traditional' method used in previous studies. First, individual clumps' paths are traced manually. For each time step at which rain appears along the path and at each position along the clumps' paths semi-automated routines measure the width and length following the same criteria specified above. The average and standard deviation of these measurements for a given path provide the clump's size (width and length) and corresponding errors. 

For all the above measurements, regardless of the instrument, clump intensities are calculated by subtracting from a given rain pixel intensity the average background intensity at the same pixel position. The average is calculated over a time interval close to that of detection and for which no rain is detected. The Doppler velocity and temperature calculation with CRISP were performed following the same technique as in \citet{Antolin_etal_2012SoPh..280..457A} and we refer the reader to that paper for details. It is important to note that with this technique upper limits to the temperatures are provided, since the micro velocities from turbulence are ignored. 

For the follow-up of individual rain features we used the CRisp SPectral EXplorer \citep[CRISPEX,][]{Vissers_Rouppe_2012ApJ...750...22V} and its auxiliary program TANAT (Timeslice ANAlysis Tool). Both are widget-based tools, which enable the easy browsing and analysis of the image and spectral data, the determination of clump paths, extraction and further analysis of space-time diagrams. Fitting piece-wise segments along the clumps trajectories in the space-time diagrams we can estimate the projected velocities (in the plane of the sky). With this method, the error (standard deviation) for projected velocities is estimated to be roughly $5~$km~s$^{-1}$. 

\section{Temperature evolution}\label{temp}

\subsection{\textit{SDO}/AIA \& \textit{SST}/CRISP observations}\label{aia_crisp}

In Fig.~\ref{fig1} and Movie~1 active region AR11903 is viewed at the East limb. The figure is a composite of three AIA filter images, combining the 193 filter (in red), dominated by the coronal \ion{Fe}{12}~193.509~\AA\ emission at $\log T=6.2$, the 171 filter (in green), dominated by the \ion{Fe}{9}~171.073~\AA\ emission at $\log T=5.9$, and the 304 filter (in blue), dominated by the transition region \ion{He}{2} 304~\AA\ emission at $\log T=4.9$. The movie clearly shows that a significant quantity of the coronal volume in the active region, filled by loops, exhibits continuous intensity variation. Such dynamical state is strongly suggestive of loops out of hydrostatic equilibrium \citep{Aschwanden_2001ApJ...550.1036A}, and possibly in a thermal non-equilibrium state. This is further supported by the fact that large parts of the coronal volume change from red to green and to blue, indicative of cooling.

During this observation the \textit{SST} was pointing at the footpoint region of these loops, as indicated in Fig.~\ref{fig1}, where a sunspot is located. The sunspot is roughly located in the middle of the FOV. 

\subsection{Cooling from EUV to H$\alpha$ temperatures}\label{euv2ha}

\begin{figure}
\begin{center}
$\begin{array}{c}
\includegraphics[scale=0.5]{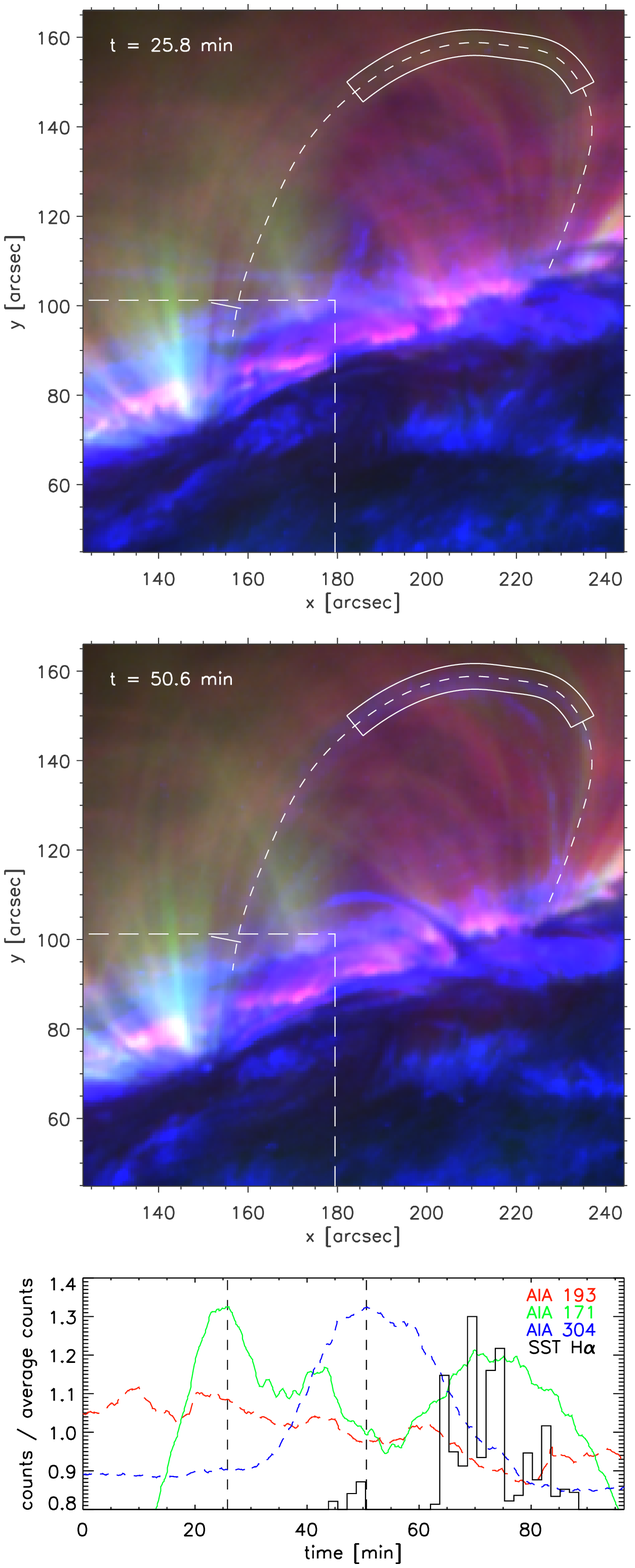} 
\end{array}$
\caption{Evolution of a loop hosting coronal rain in three AIA diagnostics and CRISP H$\alpha$ for dataset~1. The top two panels show the loop (defined as loop~2 in the text) at two instances in time (indicated by vertical dashed lines in the lower panel) in AIA 193 ({\it{red}}), 171 ({\it{green}}), and 304 ({\it{blue}}), with the loop path traced by the curved dashed line in the upper two panels. The long-dashed box outlines the extent of the \textit{SST} field-of-view, the lower left corner of which coincides with that of the panels. The lower panel shows the counts normalised to their average as function of time for AIA 193 ({\it{long-dashed red}}), 171 ({\it{solid green}}) and 304 ({\it{dashed blue}}), where these intensities have been obtained over the box around the loop top in the upper panels. For CRISP H$\alpha$ ({\it{solid black}}) we plot a histogram of intensities (scaled by 0.5 and shifted by $+0.8$ to fit nicely in the panel) that stem from measurements across the short white line perpendicular to the loop path located at about ($155\arcsec, 100\arcsec$) (which corresponds to cut C2 in Fig.~\ref{fig5}), and are weighted by the respective clumps' widths squared (see Fig.~\ref{fig7} for further details).
\label{fig3}}
\end{center}
\end{figure}

To investigate the thermal non-equilibrium scenario we have chosen a loop (hereafter loop~2, which occurs after loop~1, defined later on) in the large \textit{SDO} FOV, which exhibits this gradual change of intensity from the hot to the cool AIA channels. This loop is highlighted in Fig.~\ref{fig1} in dotted lines, it extends $\approx42$~Mm above the surface leading to an approximate length of 132~Mm for a circular geometry. Its left footpoint is also covered by the FOV of the \textit{SST}. A zoom-in figure focusing on the loop is shown in Fig.~\ref{fig3}, where two snapshots in the loop evolution can be seen in the upper two panels, separated by $\approx24$~min, while the lower panel shows the integrated intensity in different AIA channels along the top part of the loop, as well as the size-weighted coronal rain intensity in H$\alpha$ (defined later on) close to the loop footpoint. These panels show clearly a progression in the AIA  channels from 193 dominated emission to 304 dominated emission. Prior to $t\simeq$15\,min the loop top is only visible in 193 but decreasingly so between $t=10-17$\,min. From that point on 171 starts dominating clearly until peaking at $t=25.8$\,min. Only when 171 has already decreased to a local minimum plateau do we observe an increase in 304 intensity at the top, which takes over from $t=41.0$\,min and peaks at t=50.6\,min, close to 25\,min after the peak in 171 signal. The first sign of coronal rain in H$\alpha$ in this loop is already detected prior to the 304 peak.

\begin{figure*}
\begin{center}
$\begin{array}{c}
\includegraphics[scale=0.65]{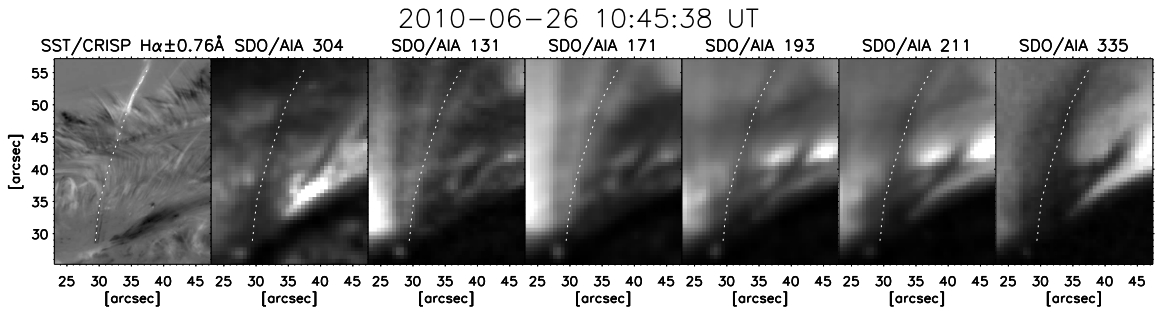} 
\end{array}$
\caption{Multi-panel figure showing a snapshot of a clump falling towards the sunspot in dataset~1. From left to right, a Doppler image from \textit{SST}/CRISP in H$\alpha\pm0.76~$\AA, \textit{SDO}/AIA 304, 171, 193, 211 and 335. The dotted lines follow the centre of the condensation, observed in emission above the spicular layer (above $y\approx47\arcsec$) and in absorption below this layer. Notice the kink in the trajectory of the clump at $y\approx35\arcsec$.
\label{fig4}}
\end{center}
\end{figure*}

\begin{figure}
\begin{center}
$\begin{array}{c}
\includegraphics[scale=0.8]{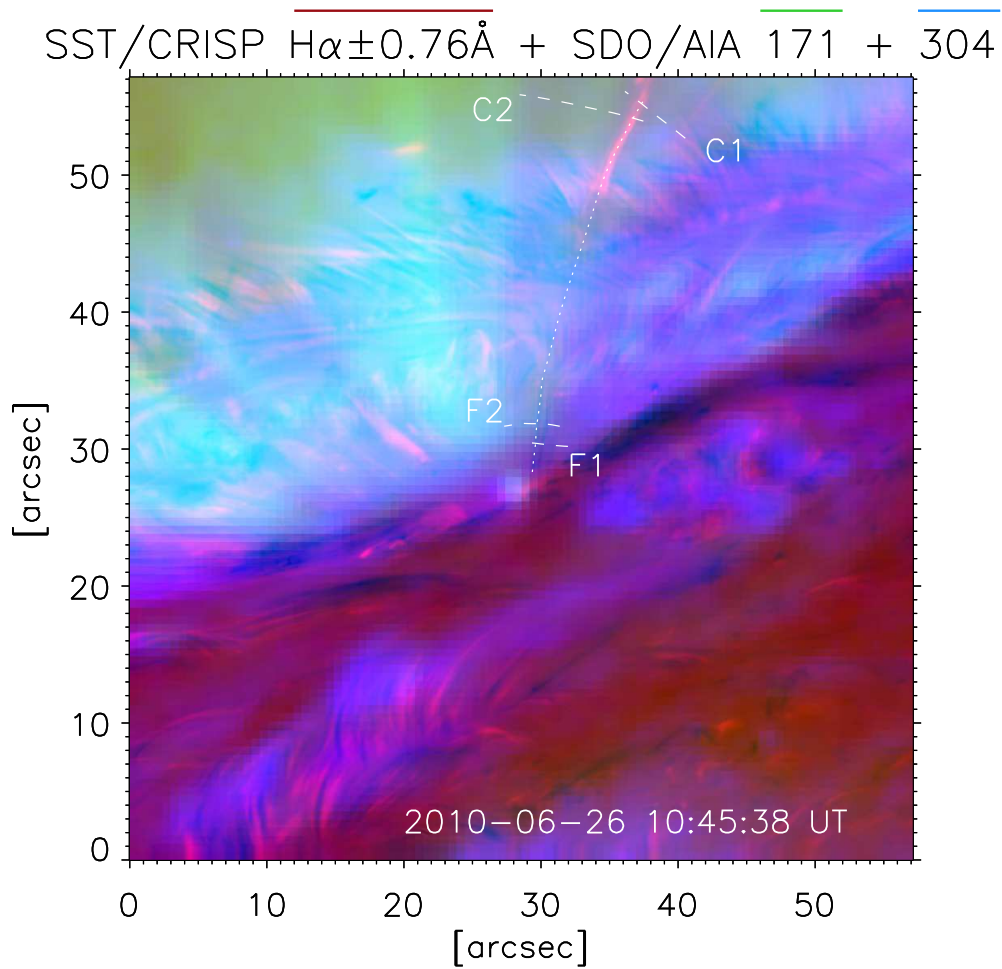}
\end{array}$
\caption{Composite image combining a Doppler image from \textit{SST}/CRISP in H$\alpha\pm0.76~$\AA\ (in \textit{red}), \textit{SDO}/AIA 171 (in \textit{green}) and \textit{SDO}/AIA 304 (in \textit{blue}) for dataset~1. The image shows the same snapshot of the falling clump as in Fig.~\ref{fig4}, whose trajectory is indicated by the dotted curve. Four transverse cuts to the main axis of two rainy loops are shown, located at two different heights: C1 and F1 for loop~1 (C2 and F2 for loop 2) at coronal and chromospheric heights, respectively. The observed clump belongs to loop 1. Movies~$3 - 6$ are zooms into this figure, showing the detected clumps crossing cuts C1, C2, F1 and F2 respectively (where the FWHM of the fit to the clumps' widths is plotted over in red).
\label{fig5}}
\end{center}
\end{figure}

While AIA 304 has a single-peaked light curve, AIA 193 and 171 show much more variability on the scale of a few to 10-15 minutes. The variations in H$\alpha$ are much more extreme and on a shorter timescale than for the AIA channels, but this is to be expected considering the difference in cadence and the fact that the AIA intensities are integrated over a large area, while the H$\alpha$ intensities stem from single rain clump measurements. A large quantity of H$\alpha$ clumps is observed falling across the \textit{SST} FOV with basically two slightly different trajectories but converging to similar locations, apparently within the umbra of the sunspot. In Fig.~\ref{fig4} an example of a large clump is shown in an H$\alpha$ Doppler image ($\pm0.76~$\AA, left panel) and in AIA filters. The track of the clump, shown in dotted lines coincides with a dark loop structure of larger width in the EUV passbands. This dark structure matches with a bright loop structure above the limb (hereafter loop~1), adjacent (to the right) to loop 2. A composite image of the \textit{SST}/CRISP FOV (Doppler image H$\alpha\pm0.76$~\AA\ in red) together with AIA 171 (in green) and AIA 304 (in blue) is shown in Fig.~\ref{fig5}, where the same rain event as in Fig.~\ref{fig4} is shown. 

\begin{figure}
\epsscale{1}
\plotone{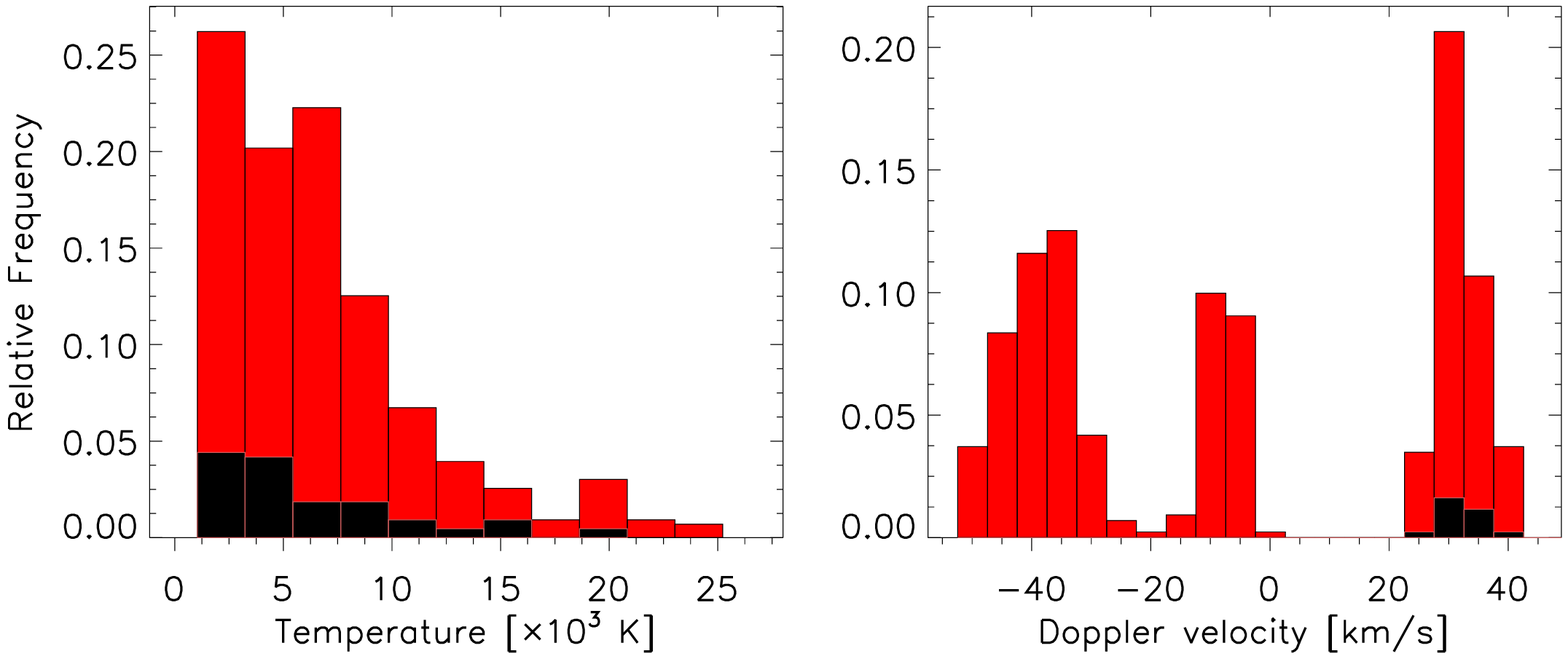}
\caption{Histograms of temperature (left) and Doppler velocity (right) for the H$\alpha$ clumps detected with \textit{SST}/CRISP in dataset~1, following the method described in Section~\ref{methods}. The black histogram corresponds to events whose measurements have a standard deviation above 10\% of the mean. Positive and negative Doppler velocities correspond, respectively, to redshifts and blueshifts.
\label{fig6}}
\end{figure}

In order to identify all the clumps, for instance those belonging to loops~1 and 2, a semi-automatic routine was written detecting all rain-like structures crossing a specific path in the FOV (see Section~\ref{methods}). This routine calculates the intensity, Doppler velocity, the width and the length for each clump. We select two cuts, C1 and C2, roughly transverse to loops~1 and 2, respectively, plotted in Fig.~\ref{fig5}, crossed by most of the clumps. Movies~3 and 4 show the detected clumps as they cross C1 and C2, respectively (in red the FWHM of the fit to the clumps' widths). These cuts can be seen to be roughly perpendicular to the trajectory of the clumps. The algorithm confidently detects 46 and 95 clumps for C1 and C2, respectively (as shown in the movie, a myriad of smaller clumps exist, but these are too faint and cannot be confidently detected by the algorithm). Despite the apparent crossing of these loops (evidenced by the intersection of C1 and C2), these are actually well separated along the LOS (the algorithm further ensures that no clumps are common to both C1 and C2). This is evidenced by the different Doppler velocities between clumps belonging to separate loops. Indeed, clumps belonging to a specific loop will share a common trajectory and therefore a similar Doppler velocity in general. This correlation is also obtained in 3D numerical simulations of thermal instability reproducing coronal rain and prominence threads \citep{Luna_etal_2012ApJ...746...30L}. Figure~\ref{fig6} shows a histogram of Doppler velocities and derived temperatures for all detected clumps. Positive and negative values correspond, respectively, to redshifts and blueshifts. We can see that the Doppler velocities concentrate in 3 different groups, one peaking at $\approx30$~km~s$^{-1}$, another at $\approx-10$~km~s$^{-1}$ and the last one at $\approx-40$~km~s$^{-1}$. The first and second groups correspond to the clumps crossing C1 and C2 respectively, while the third corresponds to the Doppler velocities of the clumps in loop 2 at chromospheric heights (cut F2, defined in Section~\ref{clumpcont}). The projected velocities of the clumps crossing C1 and C2 are found to be similar, with an average of $110$~km~s$^{-1}$. This ensures that the angles of fall for each family of clumps is different, and similar within each family (as indicated by the formulae for the falling angle calculated in Paper~1). We conclude therefore that the C1 and C2 clumps correspond to 2 different loops. 

The clumps crossing C2 are found to follow a dark structure in the EUV passbands similar to that plotted in Fig.~\ref{fig4}. This dark EUV structure appears bright above the limb and matches with loop~2. The geometry of the loop seen in Fig.~\ref{fig3} and the direction of the flow suggest that the material is coming towards the observer, matching with the negative Doppler velocities found in Fig.~\ref{fig6} for the C2 clumps. We can therefore confirm that the C2 clumps belong to loop 2. 

As shown by \citet{Anzer_Heinzel_2005ApJ...622..714A} the H$\alpha$ intensity from filaments is strongly correlated to darkening in EUV pass bands, and is due mainly to continuum absorption from neutral hydrogen, neutral helium and singly ionised helium (and to a lesser extent to volume blocking, especially in the case of small size structures as coronal rain clumps). Therefore, both the volume and the composition of the emitting structure are important factors that can be correlated to the detected intensity variations in the EUV passbands. In the lower panel of Fig.~\ref{fig3} we thus plot as a black histogram the H$\alpha$ intensity multiplied by the width squared (assuming axi-symmetry for the clump) for the clumps crossing C2. The bulk of the distribution spans over 25 minutes and the peak is shifted by $\approx20$ min from the 304 peak. However, the H$\alpha$ detection is made at the footpoints, while the AIA peaks are first observed at the apex. The clumps are observed to fall with average total speeds of 110~km~s$^{-1}$ and accelerations around $0.1$~km~s$^{-2}$. Assuming no initial velocity, constant acceleration and a circular loop, the travel time of the cooling plasma is $20\pm10$~min, which locates in time the origin of the clumps close to the 304 peak. This result supports the common origin for the H$\alpha$ clumps with the rain observed in 304 in loop 2. 

The temperature histogram in Fig.~\ref{fig6} shows that all the detected clumps have very cool temperatures, mostly below 10000~K, with a peak at $2000-5000$~K. As explained in Section~\ref{methods} these temperatures correspond to upper limits. While we postpone a discussion on these low temperature values to the pertinent section, these results clearly indicate that loops 1 and 2  undergo full catastrophic cooling to chromospheric temperatures. 

\subsection{EUV variation associated with H$\alpha$ rain emission}\label{euvassocha}

\begin{figure}
\begin{center}
$\begin{array}{c}
\includegraphics[scale=1.]{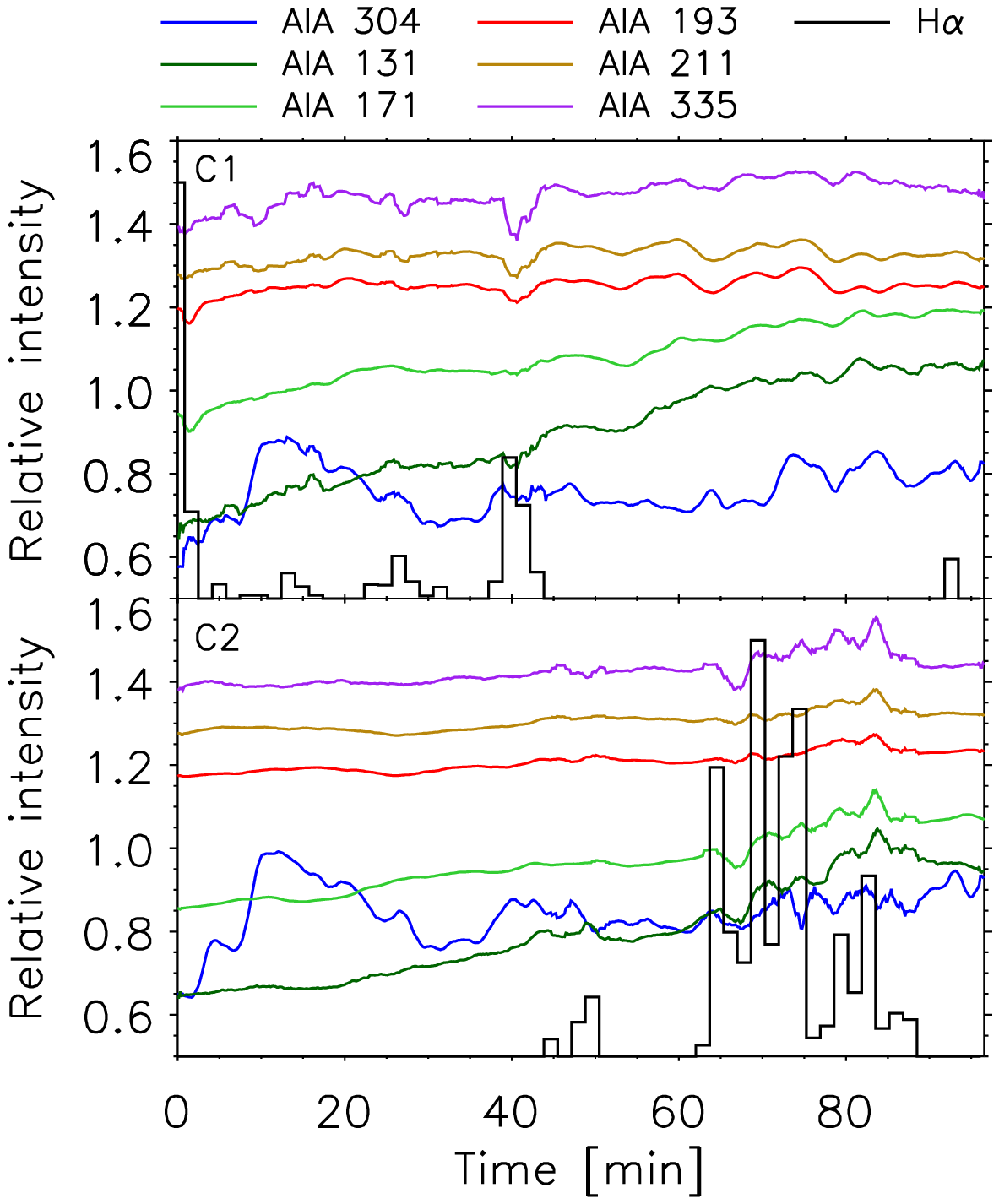} 
\end{array}$
\caption{Time evolution of the intensity across cuts C1 and C2 (corresponding to loops 1 and 2 in dataset~1, respectively) for multiple AIA filters (indicated at the top with their respective colour) and the condensations detected with CRISP in H$\alpha$ (in \textit{black}). The histogram of H$\alpha$ intensities is constructed by taking a bin size of a 100~s, and for each time bin the intensity of all contained clumps are summed, each multiplied by the clump's width squared. The AIA intensities at each time step are averages over a region of $1.8\arcsec$ in width in the transverse direction to the cuts and same length as the cut, except at times where condensations are detected in H$\alpha$. For these, the average is not calculated over the entire cut but limited to the positions along the cut where the clump is detected (defined by the width and the location of the clump).
\label{fig7}}
\end{center}
\end{figure}

It is interesting to compare the EUV and H$\alpha$ intensity variations across the C1 and C2 segments. This is shown in Fig.~\ref{fig7}. For each time step we average the intensity in each channel over a small region centred around each cut. For H$\alpha$, as in Fig.~\ref{fig3}, we plot the average of the intensity times the square of the width for each clump (see figure caption for more details). Also, for the times at which a clump is detected the intensity in each EUV channel is calculated over the location of the clump only, in order to discern better a possible effect on the EUV passband for the passing clump. For the clumps crossing C1 we can see that most of the time groups of clumps close in time produce a decrease in all of the EUV intensity channels (at times $t\approx0, 10, 27, 40$~min), except 304, for which an intensity increase is generally detected. For C2 a clear correlation is also detected. Quasi-periodic fluctuations in the EUV passbands observed in the time range t=$60-90$~min seem to match the H$\alpha$ intensity fluctuation in the same time range. Contrary to the previous case here a maximum in the H$\alpha$ intensity seems to be positively correlated with all EUV passbands. Indeed, each peak of the former seems to correspond to a local peak in the latter. 

\subsection{\textit{Hinode}/SOT, \textit{IRIS}/SJI \& \textit{SDO}/AIA observations}

We now turn to the second set of observations involving SOT, SJI and AIA. A look at Fig.~\ref{fig2} and Movie~2 immediately indicates the large presence of cool material above the active region. In red, green and blue we show \ion{Ca}{2}~H emission from SOT, \ion{Si}{4} 1400 emission from \textit{IRIS}/SJI and \ion{Fe}{9}~171 emission from AIA, respectively. We can see the presence of two kinds of cool structures above the spicular layer. Prominence material is observed flowing relatively slowly roughly horizontally, while coronal rain is observed falling relatively faster along loops bright in AIA 171 (other cool static prominence-like structures are also observed, e.g. at $(x,y)\approx(-160,970)$). Here we will concentrate on a loop exhibiting coronal rain, with one footpoint at $(x,y)\approx(-120,930)$ and the other apparently at $(x,y)\approx(-150,955)$, expanding in the corona up to a height of roughly $40\arcsec$ (leading to a length of roughly 180 Mm assuming circular geometry). The rain appears from 23:12 UT to 23:30 UT (the end of the co-observation sequence) and is observed to initiate at the apex of the loop and to fall towards the farther footpoint. 

A zoom-in onto the loop of interest is shown in various wavelengths in Fig.~\ref{fig8} and Movie~7. Contrary to the first dataset, here we are able to observe the chromospheric emission from the rain well above the spicular layer, but we are not able to follow it once it goes below this layer. 

\subsection{Cooling through transition region temperatures}\label{euv2ca}

In order to follow closely in time the evolution of temperature in each wavelength we plot in the upper panel of Fig.~\ref{fig9} the integrated emission for each filter for coronal rain pixels belonging to the previously indicated loop. As seen in Movie~2, half-way through the coronal rain event a prominence that appears to be in the background crosses the upper portion of the loop. In order to minimally reduce such LOS projection effects the selection of clump paths and the times and locations of rain along such paths is done manually and is done ensuring that the prominence is not along the LOS. In the figure we can see that the first emission of rain appears in the \textit{IRIS}/SJI images, before showing up in the \ion{Ca}{2}~H filter. The SJI 1400 filter comes first (at $t_0\approx44.5$~min), followed by a peak in SJI 1330 ($t=t_0+2.5$~min), then SJI 2796 ($t=t_0+5.5$~min) and a small peak in \ion{Ca}{2}~H roughly at the same time. The light curves generally show short-timescale variability. While all the SJI channels have a first large peak followed by gradually smaller peaks, the \ion{Ca}{2}~H emission starts with a small peak and increases over the next 5 minutes, with the highest peak at $t=t_0+8$~min. Interestingly, the time lag between the first or highest peak of each diagnostic seems to increase as the rain becomes visible in each: the \ion{C}{2} peak follows the \ion{Si}{4} peak after $\sim$2.5\,min, \ion{Mg}{2}~k follows \ion{C}{2} after $\sim$3\,min, \ion{Ca}{2}~H follows \ion{Mg}{2}~k after $\sim3-3.5\,$min. Although the dynamic range differs, part of the increases and decreases are visible accross several diagnostics. For instance, the initial \ion{C}{2} peak around 46.5\,min seems to coincide roughly with a secondary \ion{Si}{4} peak, the initial \ion{Mg}{2}~k increase and peak around 49\,min is mirrored in minor by \ion{Ca}{2}~H, while AIA 304 and 171 seem to follow each other closely throughout.

The behaviour of the AIA filters, and especially 171, is strikingly different. Movie 2 shows that most of the loop is bright and stays bright in 171, from a first peak about 10~min before first appearance of coronal rain in SJI 1400. No EUV darkening co-spatial to the rain is observed in this case, as was the case for dataset~1.  Both AIA 171 and 304 increase slightly in intensity during the rain event, with the 304 emission of the rain starting at $t=t_{0}+3.5~$min, roughly at the same time as the \ion{Mg}{2}~k emission. This is more clearly seen in Movie~7. This behaviour seems contrary to that observed in the loops in the first dataset with CRISP and AIA, in which 171 emission at the loop apex peaks either prior or following a rain event and is at a minimum during the rain event, and in which 304 peaks after 171 and roughly at the same time or prior to H$\alpha$ (cf. Fig.~\ref{fig3}). Multiple small peaks are also observed in the AIA 171 emission, in a quasi periodic fashion separated by a few minutes, similar to the behaviour observed in dataset~1, shown in Fig.~\ref{fig7}. 

Another useful way to visualise the evolution of emission is through the cumulative light curves, plotted in the lower panel of Fig.~\ref{fig9}. In this figure we can see the same progression throughout the filters, with the variability reflected now in the slope of each light curve. Accordingly, the behaviour between the SJI filters is very similar, characterised by two different slopes: an initial fast rise with a steep slope (until $t\approx50~$min) followed by a more gradual rise. The fast rise can also be seen in movies 2 and 7: the rain brightens as soon as it appears close to the apex of the loop, especially in SJI 1400 and 1330. This two step behaviour suggests two different cooling phases. The \ion{Ca}{2}~H emission appears mostly during the second cooling phase, it has roughly the same slope throughout the evolution, reflecting a similar rise and decay of the intensity in the upper panel. Accordingly, movies 2 and 7 show a more gradual emission in \ion{Ca}{2}~H. The behaviour between the AIA 304 and 171 filters is very similar and constant on average, leading to a roughly constant slope in the cumulative plot of Fig.~\ref{fig9}.

\begin{figure*}
\epsscale{1.}
\plotone{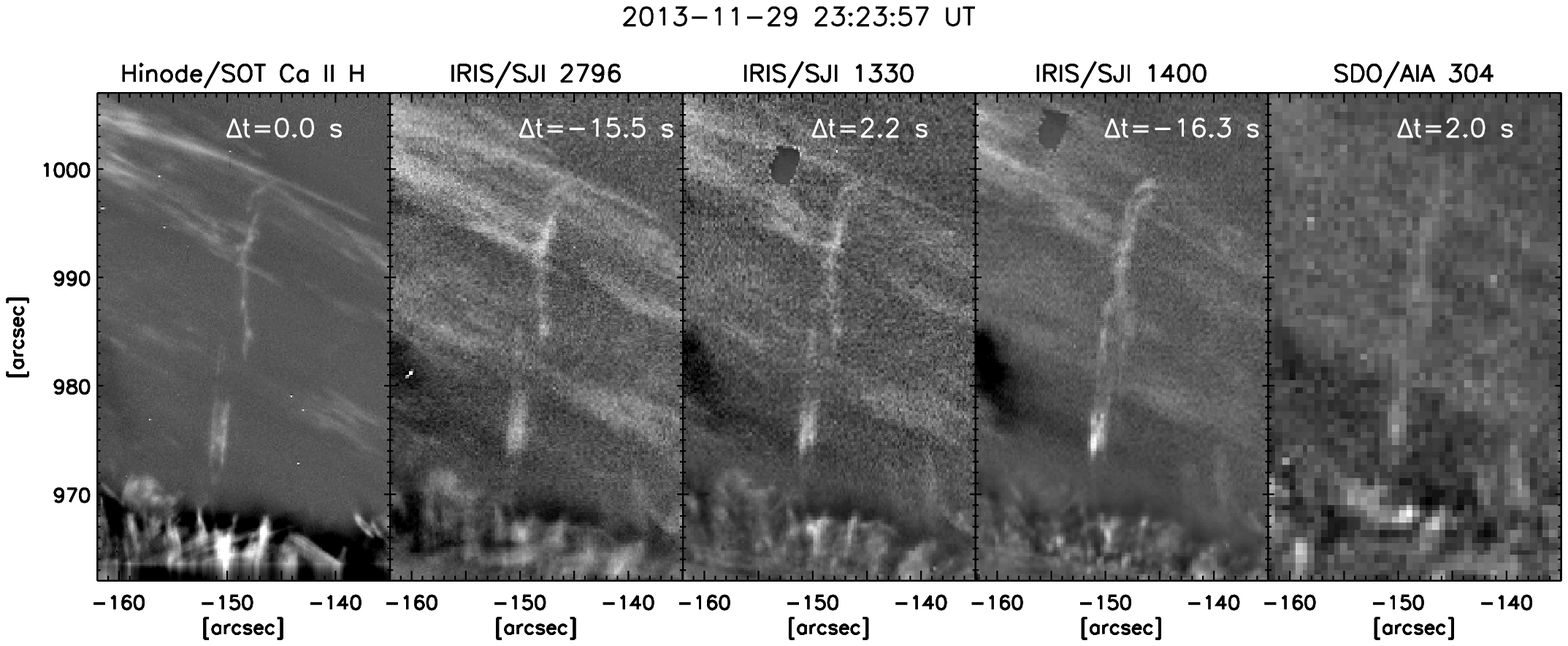}
\caption{Multi-panel showing a snapshot of the coronal rain event in dataset~2. From left to right we have \textit{Hinode}/SOT in \ion{Ca}{2}~H, \textit{IRIS}/SJI 2796, 1330, 1400 and \textit{SDO}/AIA 304. Each snapshot corresponds to the closest snapshot in time to that taken by SOT, and for each the average background is subtracted. The time difference of each snapshot with respect to that of SOT is indicated in the upper part of each panel. An animation of this figure can be found in the online material as Movie~7.
\label{fig8}}
\end{figure*}

\begin{figure}
\begin{center}
$\begin{array}{c}
\includegraphics[scale=0.7]{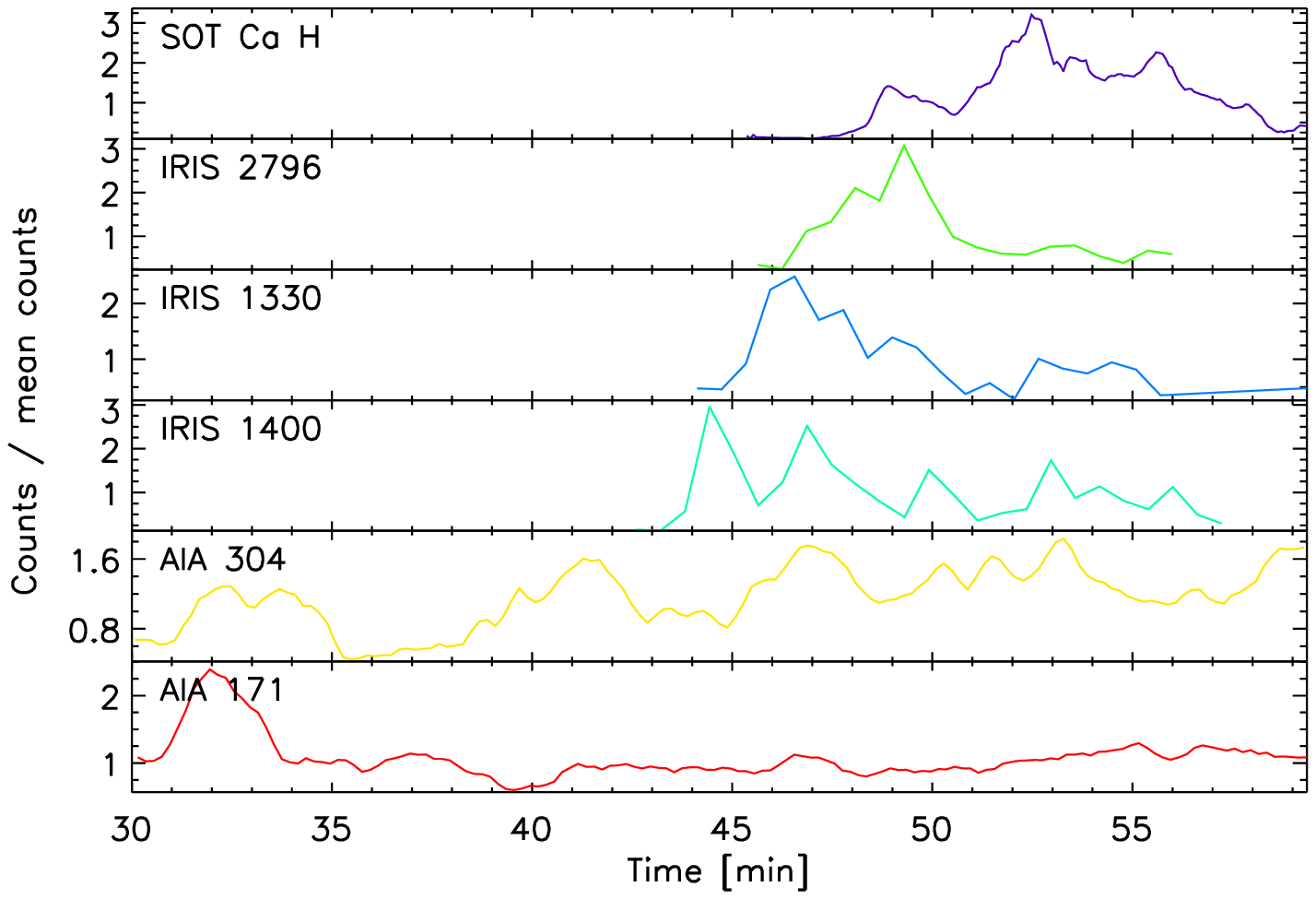}  \\
\includegraphics[scale=0.7]{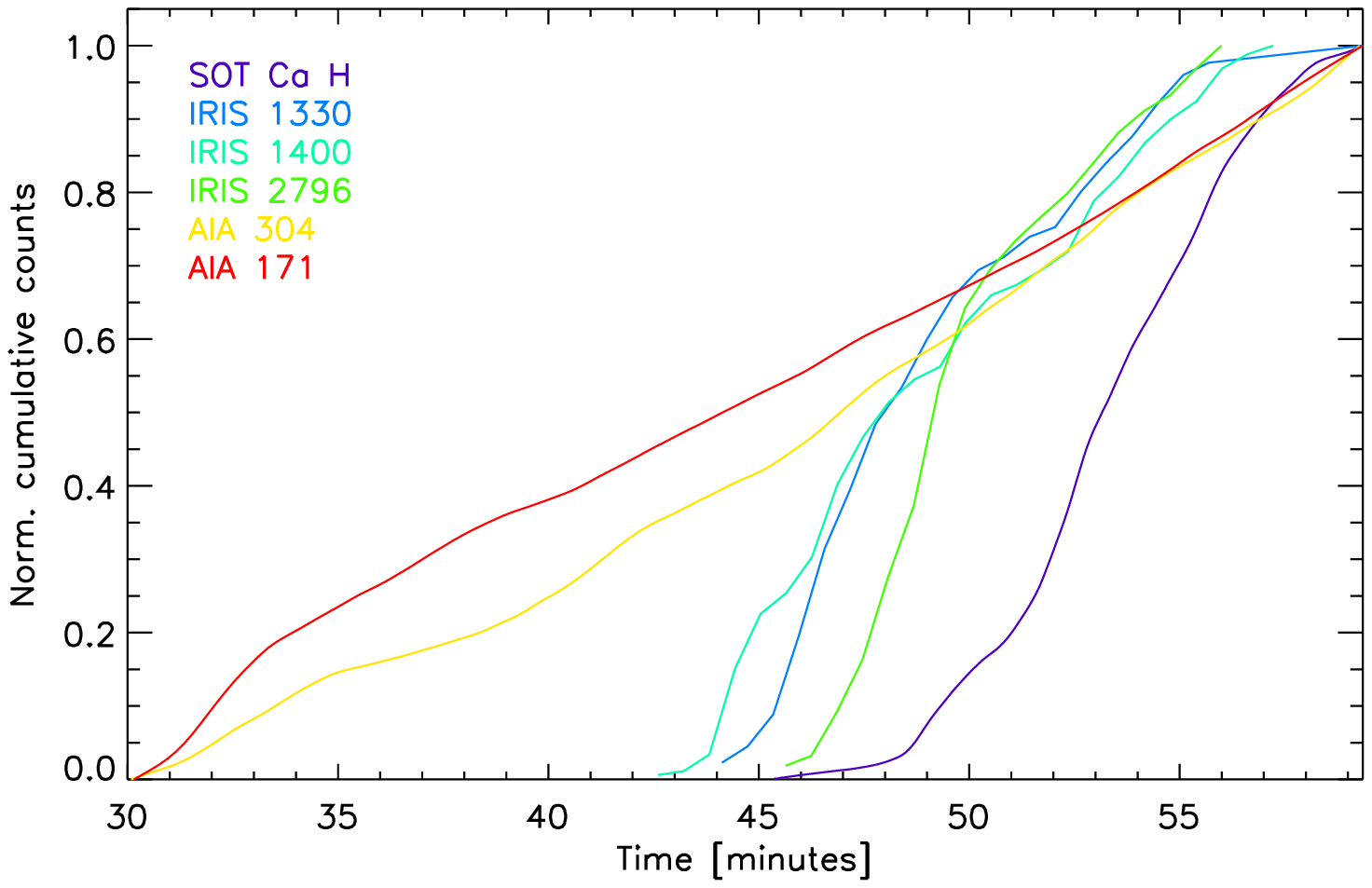}
\end{array}$
\caption{Light curves ({\it top panel}) and cumulative light curves ({\it bottom panel}) of the coronal rain pixels in the loop highlighted by the rain in Fig.~\ref{fig8} of dataset~2. In both panels the \textit{Hinode}, \textit{IRIS} and \textit{SDO} diagnostics are differentiated by colour: \ion{Ca}{2}~H ({\it purple}), \ion{C}{2}~1330 slit-jaw ({\it blue}), \ion{Si}{4}~1400 slit-jaw ({\it cyan}), \ion{Mg}{2}~k~2796 slit-jaw ({\it green}), AIA~304 ({\it yellow}), and AIA~171 ({\it red}). The SOT and AIA light curves have been smoothed over 7 and 3 time steps, respectively, close to the cadence of the \textit{IRIS} slit-jaw images.
\label{fig9}}
\end{center}
\end{figure}

Figs.~\ref{fig8}, \ref{fig9} and Movie~7 further show that the emission in all filters co-exists in the loop during most of the catastrophic cooling event, but especially during the second phase of slow cooling. This strongly suggests a multi-temperature structure for the thermally unstable loop. This is further supported by the increase of the AIA 171 intensity along the loop during the rain event. This will be further discussed in Section~\ref{disc_multi}.

The apparent mismatch with the AIA 304 channel, this one occurring 3.5~min after first occurrence in SJI 1400, could be due to the low sensitivity and lack of resolution of AIA with respect to the other instruments. A small clump will not become bright enough above the noise level unless it either increases in intensity or becomes larger. In Movie~7 we can see that AIA 304 emission is barely visible above the noise level, supporting the interpretation.

\section{Morphology}\label{morph}

\subsection{Rain showers with cool cores and associated continuous EUV darkenings}

Figure~\ref{fig3} and Movie~1 show that the downward flow along many of the loops in the active region seen in AIA 304 has a continuous character with little clear substructure, especially in the direction of motion. This is also observed in the lower part of the loops shown in Fig.~\ref{fig4}, where the downflow appears dark in absorption. However, as revealed by movies 3 and 4, the falling H$\alpha$ clumps observed with CRISP come in various sizes and intensities along the same paths. Fig.~\ref{fig5} shows an example of a clump in H$\alpha$ in emission above the spicular layer, and extending onto the disc towards the umbral region of the sunspot (marked by a dotted line in the figure). The on-disc part of the elongated rain clump is observed as a thin absorption feature and is adjacent to other elongated rain clumps (also seen in Fig.~\ref{fig4}). Surrounding the H$\alpha$ emission, EUV darkening can be observed all along the clump trajectory in the lower part of the loop.  This low EUV emission is more persistent, as shown in Fig.~\ref{fig7}, thus matching the continuous (as opposed to clumpy) character observed in AIA 304 in the larger FOV of Fig.~\ref{fig1}. However, as shown in Fig.~\ref{fig7} and described in Section~\ref{euvassocha}, a close look at this low emission reveals small EUV variations, which appear correlated to the H$\alpha$ intensities. This occurs especially when several H$\alpha$ clumps appear close in time. Such groups composed of a large number of small H$\alpha$ clumps were denoted in Paper~1 as `showers'. As predicted in that work here we show that such showers, and also a small minority of large and dense enough H$\alpha$ clumps, are associated with absorption features in EUV (and corresponding intensity variations within the dark structure). Figure~\ref{fig4} shows that these absorption features are significantly wider than the H$\alpha$ emission (on the order of a few times the H$\alpha$ width), suggesting a wide transition from cooler to hotter temperatures. Also, a comparison between movies 1 and 3 (or 4) indicate that the EUV absorption features are significantly longer than the H$\alpha$ clumps (except for the few very long clumps that extend beyond the FOV, for which we cannot confirm). These results may suggest at first a scenario in which cool chromospheric cores are surrounded by warmer diffuse shells. However, the large difference in spatial resolution between CRISP and AIA may be the main factor behind this picture.

\subsection{Co-spatiality of emission and substructure}\label{cospatial}

\begin{figure}
\epsscale{0.8}
\plotone{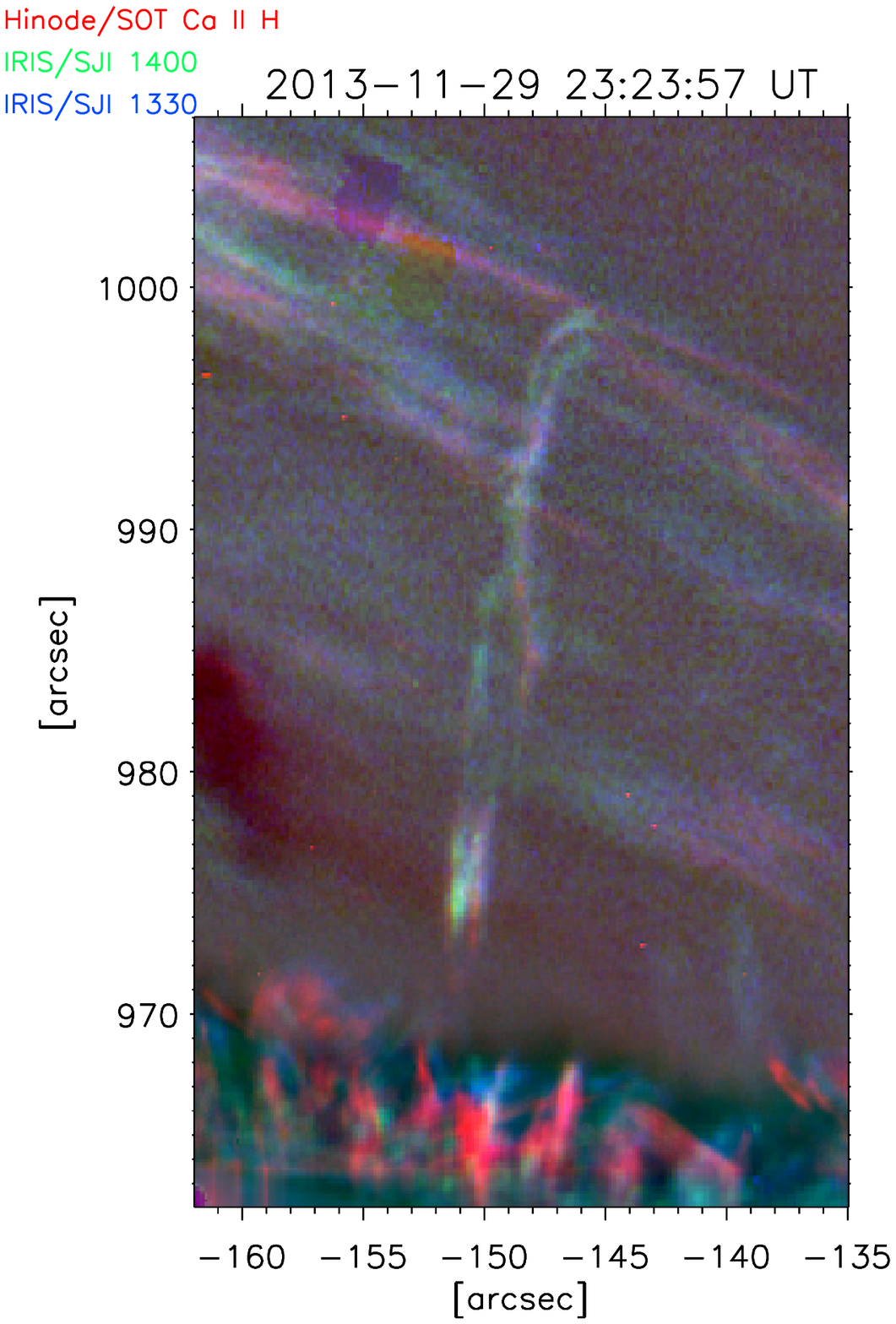}
\caption{Composite image of the rain event in dataset~2 combining \textit{Hinode}/SOT in \ion{Ca}{2}~H (in \textit{red}), \textit{IRIS}/SJI 1400 (in \textit{green}) and \textit{IRIS}/SJI 1330 (in \textit{blue}). An animation of this figure can be found in the online material as Movie~8. 
\label{fig10}}
\end{figure}

The previous scenario is further supported by dataset~2. As shown by Fig.~\ref{fig8} and Movie~7, the downflow in AIA 304 appears mostly structure-less, although with a slightly brighter head (which also appears bright in the other chromospheric and transition region filters). On the other hand, substructure can clearly be seen not only in the chromospheric filters (\ion{Ca}{2}~H and SJI 2796) but also in the transition region filters (SJI 1330 and SJI 1400). Close inspection further indicates a very similar rain structure suggesting a high degree of co-spatial emission in both chromospheric and transition region filters. This can more clearly be seen in Fig.~\ref{fig10}, which shows a composite image of \ion{Ca}{2}~H (in red), SJI 1330 (in blue) and SJI 1400 (in green). This result suggests that the lack of structure in AIA 304 may not be due mainly to a difference in temperature but rather to a lack of spatial resolution. 

\begin{figure}
\epsscale{1.}
\plotone{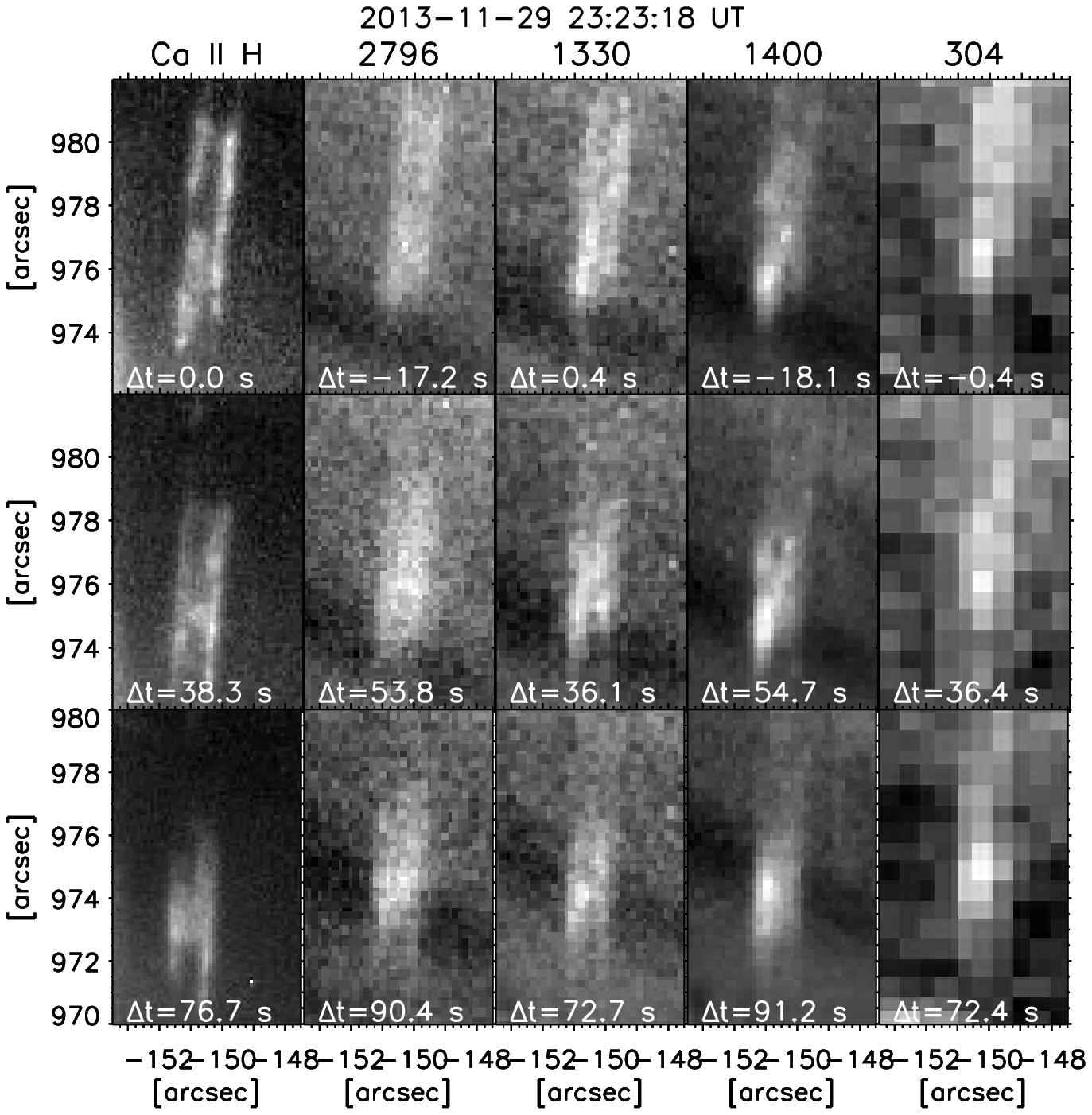}
\caption{Substructure for a rain shower in the rain event of dataset~2. The shower corresponds to the head of the rain and is shown at three different times separated by roughly 38~s from each other (shown in the rows) and for different wavelengths (shown in the columns). From left to right we have \textit{Hinode}/SOT in \ion{Ca}{2}~H, \textit{IRIS}/SJI 2796, 1330, 1400 and \textit{SDO}/AIA 304. The SJI and AIA snapshots correspond to those closest in time to the SOT snapshot in the first column of each row. The time difference with respect to the first snapshot (top left panel) is indicated in the bottom of each panel. 
\label{fig11}}
\end{figure}

To analyse more closely the degree of co-spatiality we zoom-in into the region and follow a small shower of clumps (the head of the rain, which appears bright in Fig.~\ref{fig8}). This is shown in Fig.~\ref{fig11}, where the same shower is shown at three different times along its fall, separated by $\approx40~$s from each other. The shower's length becomes shorter as it falls (passing from $8\arcsec$ to $4\arcsec$). Its width on the other hand remains mostly unchanged. However, the width of the shower (and the amount of substructure) depends strongly on the wavelength. While in AIA~304 the shower appears as a single entity, the presence of substructure is obvious in \ion{Ca}{2}~H. The shower can be seen to be composed of two strands, each one presenting inhomogeneities along their lengths. The substructure is far less evident in SJI 2796, probably due to the significant increase of opacity in this line. However, it resurfaces in SJI 1330 and SJI 1400 with similar shapes, although less clear due to the lower resolution. This picture not only proves a multi-temperature scenario for coronal rain but also proves the existence of substructure in thermally unstable loops in chromospheric to at least transition region temperatures. A look at AIA~171 (Movie~2) further shows that coronal temperatures are present throughout the loop, especially in the wake of the rain head.

\subsection{Density for a thick clump}\label{denseblob}

\begin{figure}
\begin{center}
$\begin{array}{c}
\includegraphics[scale=1.]{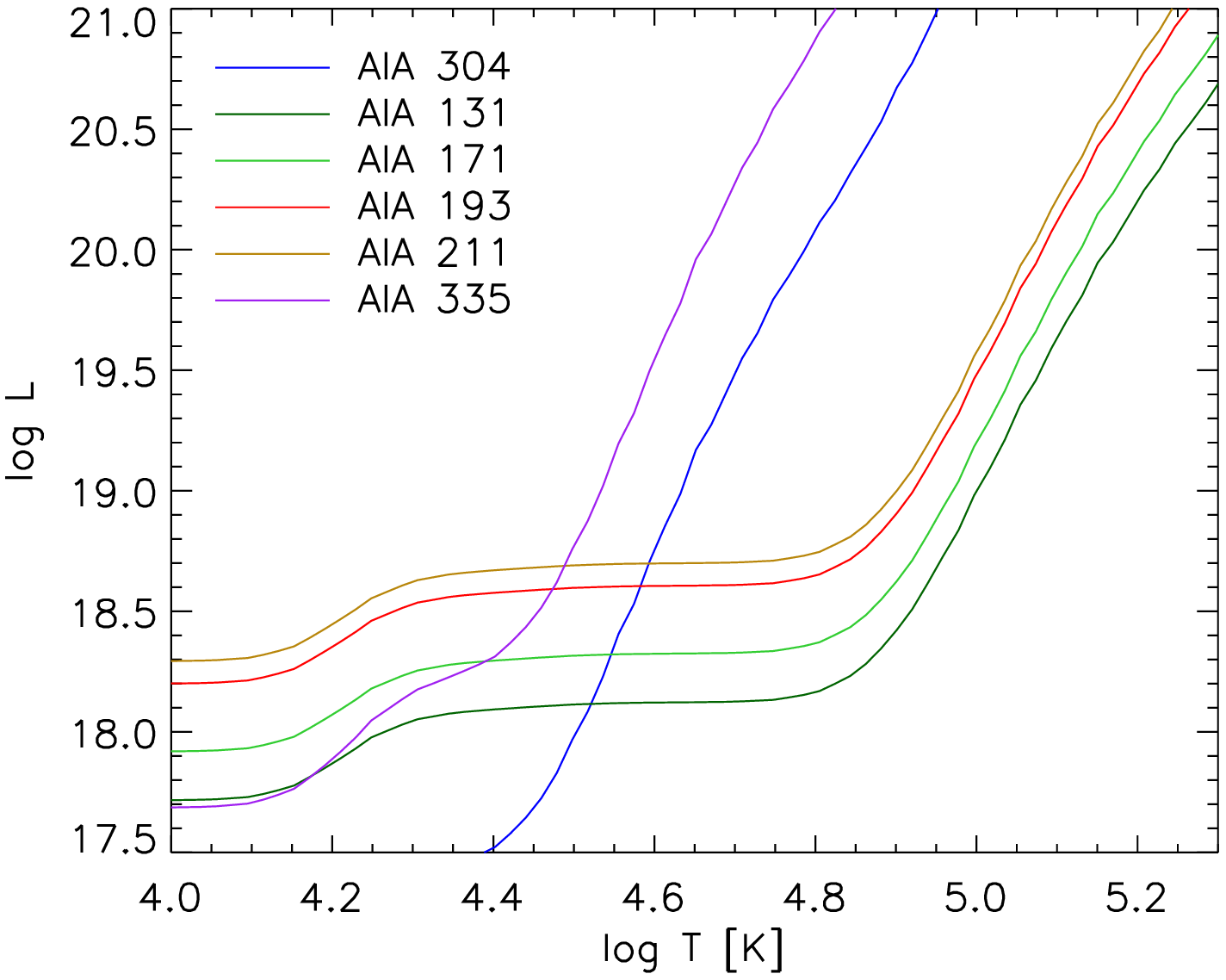} 
\end{array}$
\caption{The $L$ function (in logarithmic scale), as defined in \citet{Landi_2013ApJ...772...71L}, with respect to temperature, for the various AIA filters calculated over the trajectory of a rain condensation producing EUV darkening. 
\label{fig12}}
\end{center}
\end{figure}

Through the change in AIA emission produced by EUV absorption from hydrogen and helium in the clumps in dataset~1 we can estimate their densities following the technique by \citet{Landi_2013ApJ...772...71L}. The application of this technique is based on certain assumptions. Mainly, the plasma needs to be in ionisation equilibrium, the produced  absorption needs to be homogeneous along the LOS, and the emission in the corresponding wavelengths of the absorption needs to be minimal. While these may be questionable in the present case (a catastrophically cooling plasma could be out of ionisation equilibrium if the cooling timescale is shorter than the ionisation and recombination timescales; also, as shown previously, we have a strong degree of density inhomogeneity within the rain) the results of the applied technique can nonetheless provide important density constraints and clues on the state of the plasma. Here we do this for a large clump producing enhanced EUV darkening, similar to that shown in Figs.~\ref{fig4} and \ref{fig5}, but we postpone a full investigation on densities to future work. In Fig.~\ref{fig12} we plot the $L$ function, defined in \citet[][formula 9]{Landi_2013ApJ...772...71L}, which, at temperature of absorption is equal to the hydrogen column density. Our case is similar to their special case 2 (Section 3.2 in their paper). The background emission is measured at the same locations where the EUV darkenings are observed but at times where no H$\alpha$ clumps are observed (we take a few time steps prior and after the passage of the clump, and then take the average of the two). The EUV intensities are taken over the trajectory of the clump, seen as darkened streaks in an $x-t$ plot. In the calculation of $L$ we further assume a helium abundance of 5\%. In Fig.~\ref{fig12} we can see that the region of intersection of all curves is concentrated in $\log T_{abs}=4.4-4.6$ and $\log L=18.1-18.7$~cm$^{-2}$ (where $T_{abs}$ is the temperature of maximum absorption). Assuming axi-symmetry along the direction of propagation the LOS depth across the clump is then roughly equal to the clump's width. The clump's width in the EUV filters is $2.8\arcsec$, although it cannot be determined with precision due to the poor spatial resolution of AIA. The clump in H$\alpha$ has a maximum width of $0.6\arcsec$ (with an average of $0.4\arcsec$). From the results of dataset~2, it is likely that the larger widths in EUV are mostly an effect of the lack of resolution (which would smear out the width a factor proportional to the AIA's PSF). We therefore assume an EUV width of $1.5 - 2$ times larger than the H$\alpha$ width, leading to a width of roughly 700~km. This leads to an electron density of $1.8-7.1\times10^{10}$~cm$^{-3}$, where the scatter is most likely due to the presence of inhomogeneities within the clump. This is supported by the fact that the spectral width of the clump in H$\alpha$ indicates a temperature of $5500\pm500$~K, suggesting a multi-temperature structure, and possibly a higher density. Assuming a constant pressure within the clump the core density can then be up to $2.5\times10^{11}$~cm$^{-3}$.

\subsection{Clumpy vs.~continuous}\label{clumpcont}

As the clumps fall further down towards the sunspot in dataset~1 they become absorption features in the wing of H$\alpha$ while they gradually become bright towards line centre (see Figs.~\ref{fig4} and \ref{fig5}). These distinctive features allows us to trace them down to chromospheric levels, just before impact. Similarly as previously done for coronal heights we define two cuts at chromospheric heights roughly perpendicular to the clumps trajectories, F1 and F2, corresponding to loops 1 and 2, respectively. Both C \& F pairs are plotted in dashed curves in Fig.~\ref{fig5}. As movies 3 and 4 did for C1 and C2, movies 5 and 6 show the detection of rain clumps across F1 and F2, respectively (in red the FWHM of the fit to the clump's width). These movies indicate that the C1 clumps fall either into the umbra or towards the closer edge of the sunspot (projection effects do not allow us to firmly pinpoint the location of the fall) and the C2 clumps seem to fall mainly towards the far edge of the sunspot. Tracing of coronal rain clumps to such low heights has never been achieved before and is possible here thanks to the high falling speeds of the clumps (which produces the dark absorption features in the wing of H$\alpha$) and to the high spatial and temporal resolution of CRISP. Indeed, the total velocities (taking into account projected and Doppler velocities) vary between 80~km~s$^{-1}$ and 120~km~s$^{-1}$ for both loops. 

Due to the increase of projection effects at low heights, the high speed of the clumps and the seeing effects (which not always provides clear subsequent images), it is not possible in most cases to specify the start and end for a clump at low heights, and therefore provide a robust determination of clump lengths. It is therefore possible that two subsequent detections at the same location correspond to the same clump. Also, it is highly likely that many of the clumps detected at heights C are detected again at heights F. However, since we are more interested in regarding the rain as a flow rather than a discrete set of condensations (especially because their morphology can change dramatically as they fall), for our statistical analysis we consider that all clumps are different (and for the lengths we give an approximate value determined visually). Since there are many clumps that fail to be detected by the algorithm (due to the previously stated reasons), we consider that this procedure leads to a better approximation to the real rain population.

\begin{figure}
\begin{center}
$\begin{array}{c}
\includegraphics[scale=1]{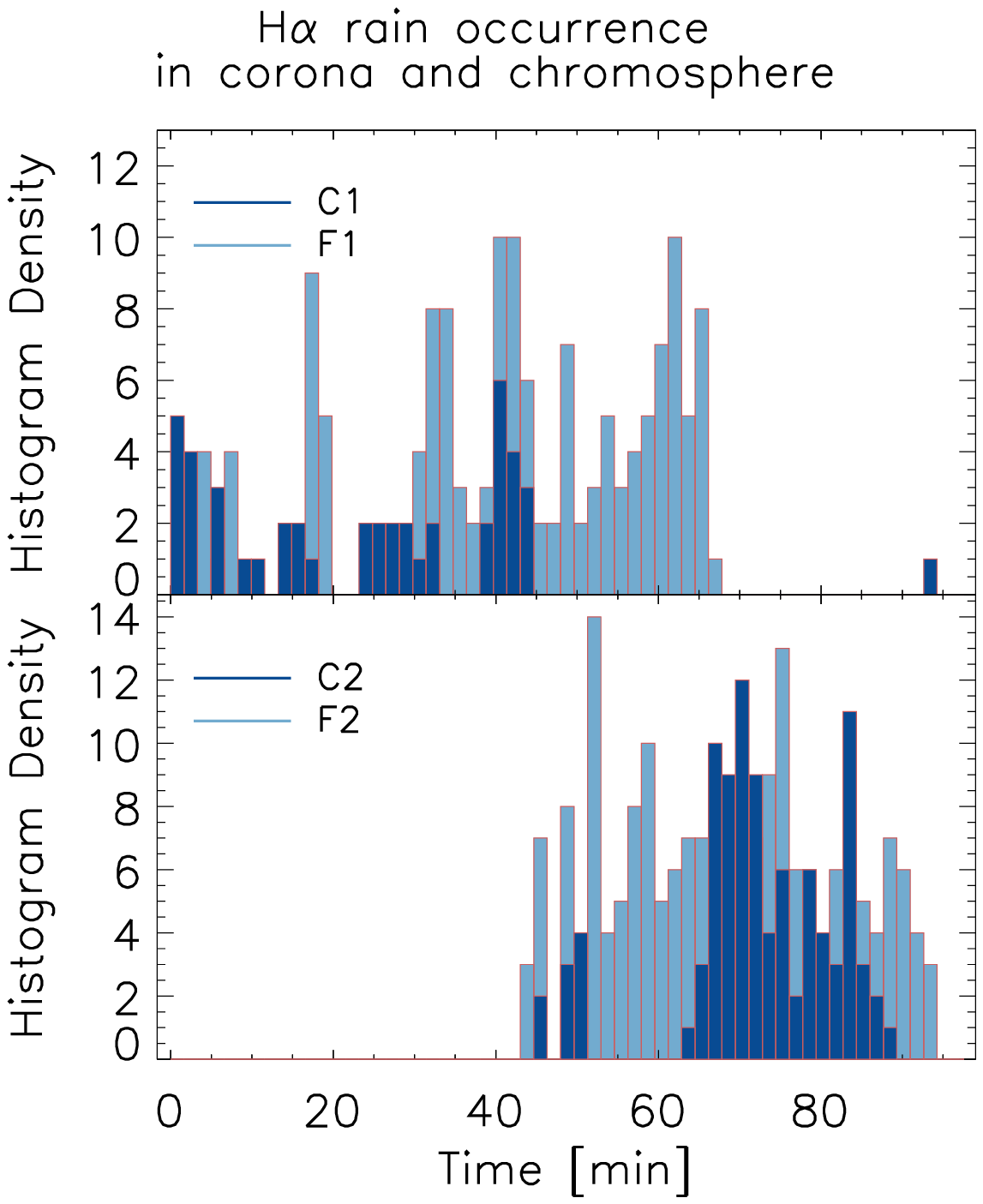} 
\end{array}$
\caption{Histograms of rain condensations detected in H$\alpha$ with \textit{SST}/CRISP for dataset~1 along cuts C1, F1 and C2, F2 for loops 1 and 2, respectively. The condensations detected at coronal heights (C cuts) constitute the deep blue histogram, while those detected at chromospheric heights (F cuts) constitute the light blue histogram. The histograms are constructed by taking a bin size of a 100~s, and for each time bin the intensity of all contained clumps are summed, each multiplied by the clump's width squared.
\label{fig13}}
\end{center}
\end{figure}

A striking feature shown by movies 5 and 6 is that at such low chromospheric heights the clumps appear significantly less clumpy (their lengths increase) and the downward flow becomes continuous and rather persistent. In order to  quantify this effect better we plot in Fig.~\ref{fig13} a histogram of clump detection weighted with their respective widths squared at the two previously defined heights for loops 1 and 2. Not only the clumps in the corona are detected at later times in the chromosphere, but we also detect many more rain crossings at those lower heights, indicating an increase in the rain flow with decreasing height (46 and 145 clumps for C1 and F1, respectively; 95 and 193 clumps for C2 and F2, respectively). In Section~\ref{disc_pers} we discuss possible reasons for this increase in clump occurrence at low chromospheric heights. 

\subsection{Strand-like structure}\label{morph_str}

\begin{figure}
\begin{center}
$\begin{array}{c}
\includegraphics[scale=1]{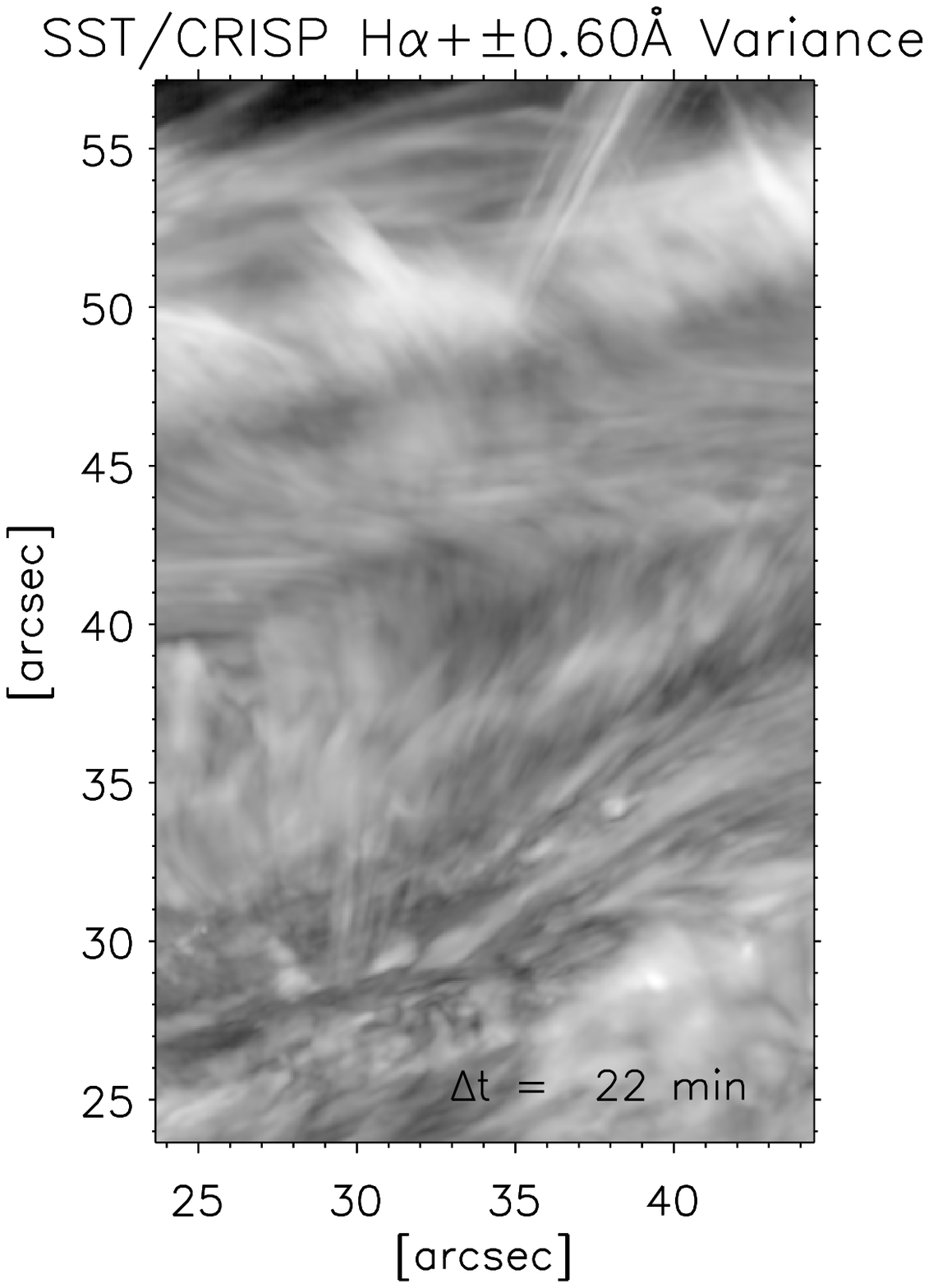} 
\end{array}$
\caption{Variance of a Doppler image in H$\alpha\pm0.6~$\AA\ with \textit{SST}/CRISP over a time of 22~min for loop 1. Various strands are visible from $50\arcsec$to $55\arcsec$ and $28\arcsec$ to $35\arcsec$ in the vertical axis.
\label{fig14}}
\end{center}
\end{figure}

As shown in Section~\ref{cospatial}, the presence of substructure in coronal rain at high resolution is not only clear at chromospheric temperatures such as those of \ion{Ca}{2}~H and SJI 2796 but also at higher  temperatures, such as those of SJI 1330 and SJI 1400. As shown in Fig.~\ref{fig11}, as the rain falls substructure is observed to be organised along strands of various lengths but rather similar widths. This is also observed in Movie~7 for the rest of the rain in this event and in dataset~1 at even higher resolution. Indeed, at the highest resolution achieved in this work (in H$\alpha$ with CRISP) similar strand-like structure appears, as shown in Fig.~\ref{fig4} and movies $3-6$. Due to the very dynamical nature of coronal rain it is not easy to discern the strand-like structure traced in time by the rain. For this, we plot in Fig.~\ref{fig14} the time variance in a Doppler image in H$\alpha\pm0.6~$\AA\ over a time period of 22 min for loop~1 and its surroundings (where the time variance is defined as the sum of the squared average-subtracted intensity for each pixel). Thanks to the emission features of the rain in the wing of H$\alpha$ both above the spicular layer and at chromospheric levels, the clumps produce bright traces in the variance image. These traces appear as various strands at both of these heights (in the figure, these are visible from $50\arcsec$to $55\arcsec$ and $28\arcsec$ to $35\arcsec$ in the vertical axis). In the figure, 6 to 8 strands with widths around $0.2\arcsec-0.4\arcsec$ can be discerned at both heights, although a larger number may exist due to projection effects. It is worth noting that the 22 min period over which the variance is taken is a significantly long time for coronal rain. Indeed, the number of clumps detected over this time period is on the order of $60-70$. Also, as calculated previously in Section~\ref{euv2ha}, the travel time for rain in these loops is in the order of $20\pm10$~min. Lastly, numerical simulations indicate that the catastrophic cooling part of limit cycles is generally short, under an hour or so \citep{Antolin_2010ApJ...716..154A,Mendozabriceno_2005ApJ...624.1080M,Susino_2010ApJ...709..499S} (although this depends on various parameters of the heating). Therefore a significant part of the catastrophic cooling for this loop may be captured in the time variance, strongly suggesting that the clumps follow well-defined fundamental magnetic structures in the corona. 

\begin{figure}
\begin{center}
$\begin{array}{c@{\hspace{-0.2in}}c@{\hspace{-0.2in}}c}
\includegraphics[scale=0.6]{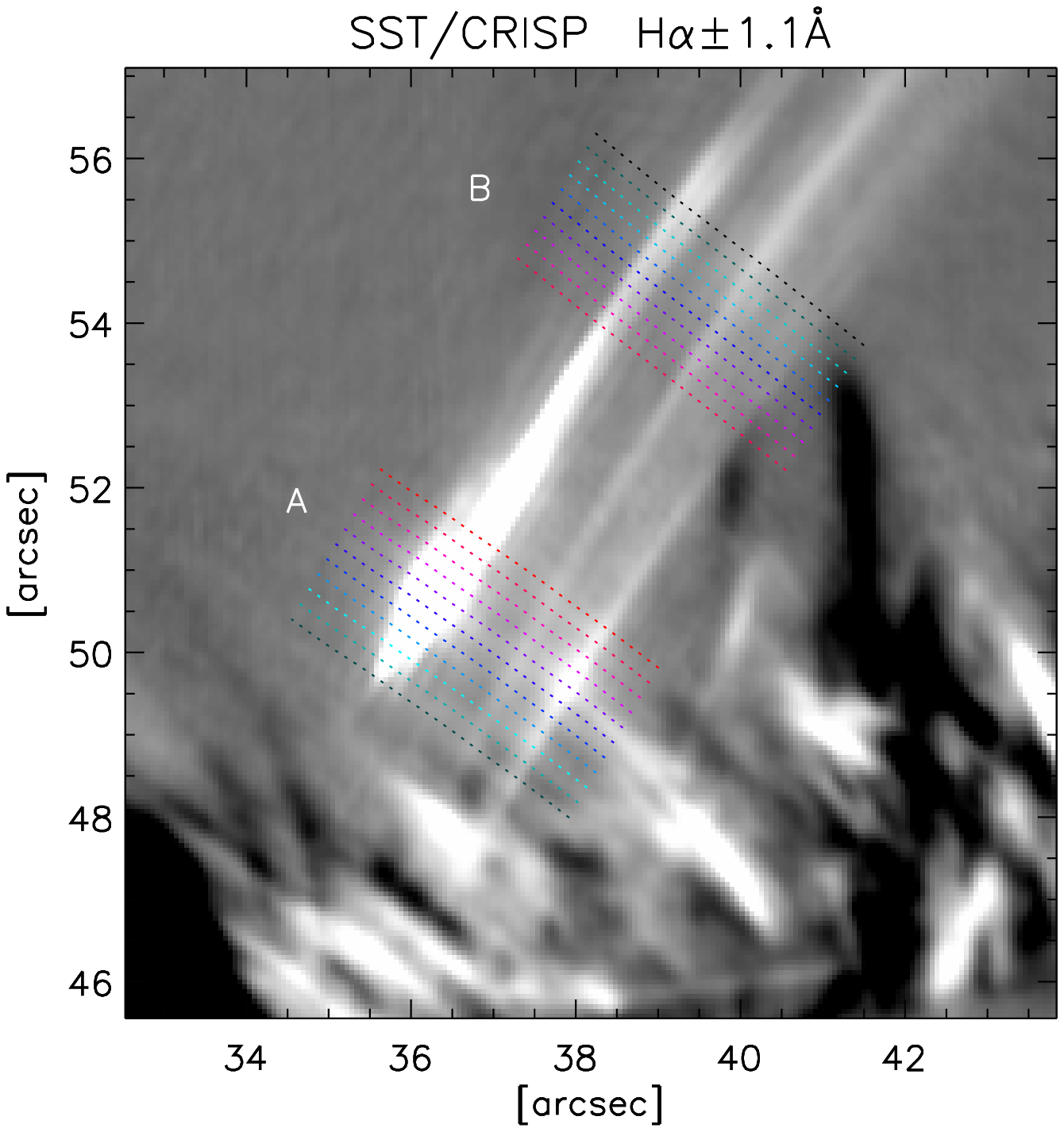}  &  
\includegraphics[scale=0.6]{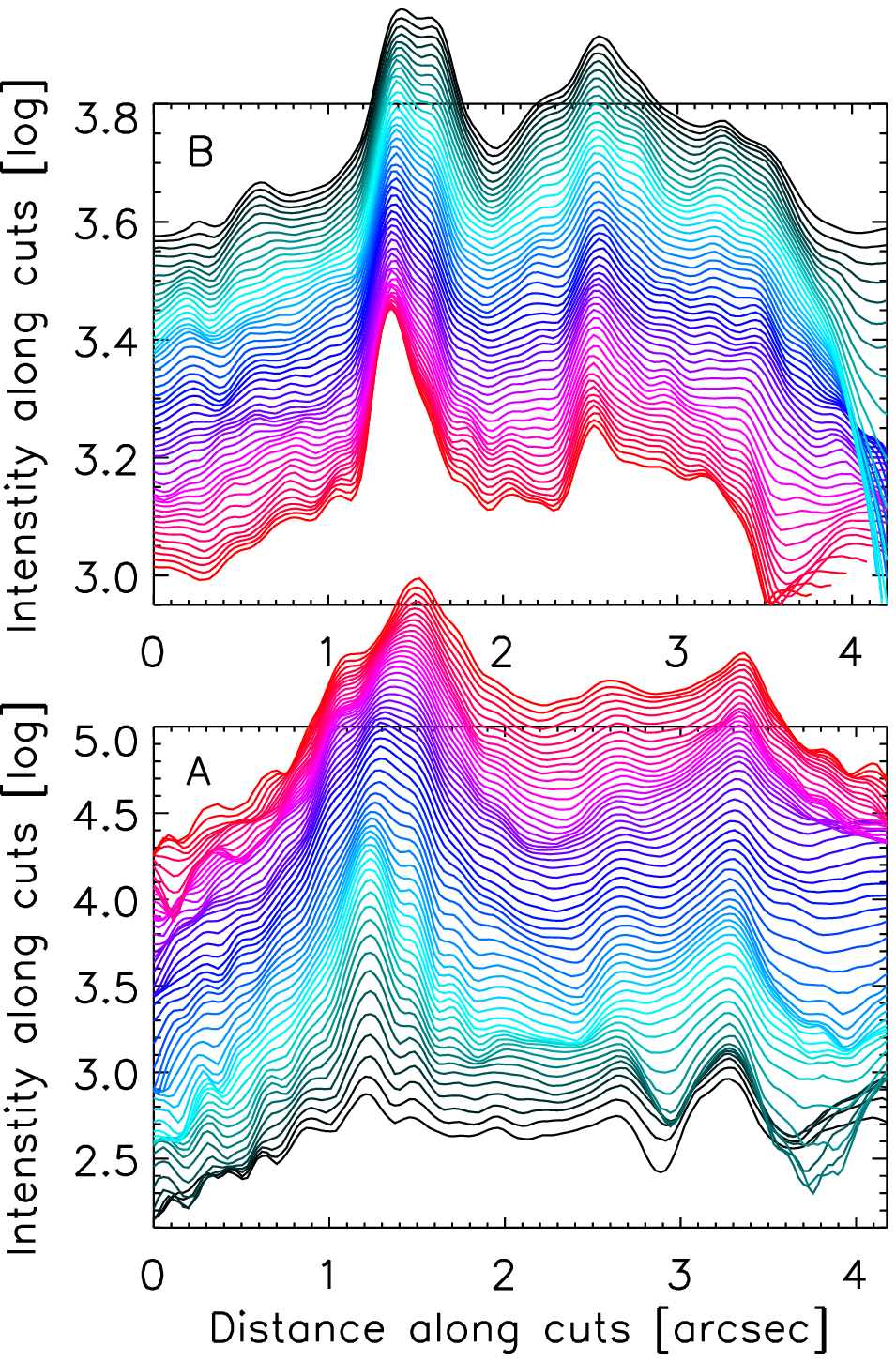} \\
\end{array}$
\caption{A rain condensation captured in H$\alpha$ by \textit{SST}/CRISP, here shown as a Doppler image (\textit{left panel}) at $\pm1.1$~\AA. The intensity of the image is saturated in order to better visualise the small dim structure parallel to the main condensation. Transverse cuts to the direction of propagation of the condensation are shown towards the head (location A) and tail (location B). The intensity along these cuts is shown in the \textit{right panels} at both A and B locations as stack curves separated by an ad-hoc value. The colour of the intensity curve corresponds to the cut of the same colour in the left panel (for location A or B).
\label{fig15}}
\end{center}
\end{figure}

Interestingly, apart from the prominent clumps as those in Fig.~\ref{fig11}, the surrounding diffuse atmosphere in dataset~1, dimly bright in H$\alpha$ also appears structured in the same way. This is especially clear at times of large clumps. An example of this is shown in Fig.~\ref{fig15}, where a snapshot of a falling clump can be seen in H$\alpha\pm1.1~$\AA. A bright clump with a long tail can be seen and, particularly, multiple strand-like structures next to it, as ripples with similar widths down to $0.15\arcsec$ or so, uniformly distributed within $4\arcsec$ of the clump. This structure can still be discerned (although with decreasing intensity) when shifting the wavelength position over a range of $\pm0.5~\AA\,$ around $1.1~$\AA, ruling out possible instrumental effects from the narrowband CRISP wavelength channels. In order to show the strand structure better we make transverse cuts across the clump at two different locations, one at the head and one at the tail of the clump. The H$\alpha$ intensity along these cuts is shown in the right panels of the figure. Next to the large peak corresponding to the bright clump, around 8 ripples can be distinguished extending significantly along the direction of propagation. This structure is highly reminiscent of the MHD thermal mode (also known as the entropy mode in the absence of thermal conduction), first predicted by \citet{Field_1965ApJ...142..531F} and later developed by \citet{VanderLinden_1991SoPh..131...79V,VanderLinden_1991SoPh..134..247V,Goedbloed_Poedts_2004prma.book.....G,Murawski_etal_2011AA...533A..18M}. Indeed, the spatial distribution of such mode is that of a main elongated dense clump with multiple smaller clumps side by side of similar widths. The generation of this wave is guaranteed by the small but non-zero perpendicular thermal conduction in the corona. Although static in an ideal scenario, this wave is expected to move together with the flow. In this scenario, the observed strand-like structure at the smallest detected scales could be the result of MHD thermal modes produced from thermal instability.

\subsection{Sizes}\label{sizes}

We now turn to the statistical determination of sizes. The methods used in the widths and lengths calculations are explained in Section~\ref{methods}. The routines used for this detection are semi-automatic. The detection of a clump is rather strict, and is based on several conditions which rule out most of the fainter and smaller clumps at diffraction limit resolution. For instance, the smaller widths that can be discerned in the cross profiles in Fig.~\ref{fig15} are mostly left out by the algorithm, in favour of precision and reduction of errors. 

\begin{figure}
\begin{center}
$\begin{array}{c}
\includegraphics[scale=0.7]{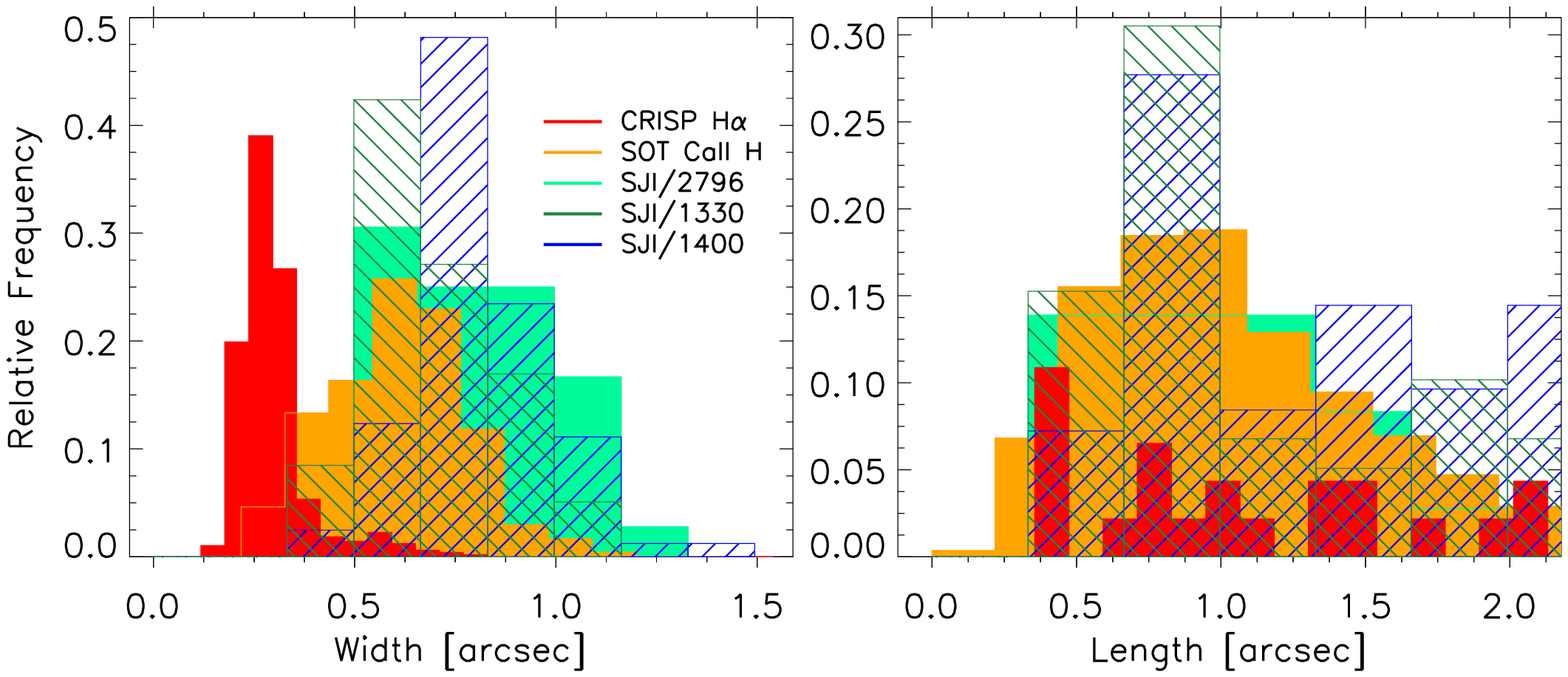}  \\
\includegraphics[scale=0.7]{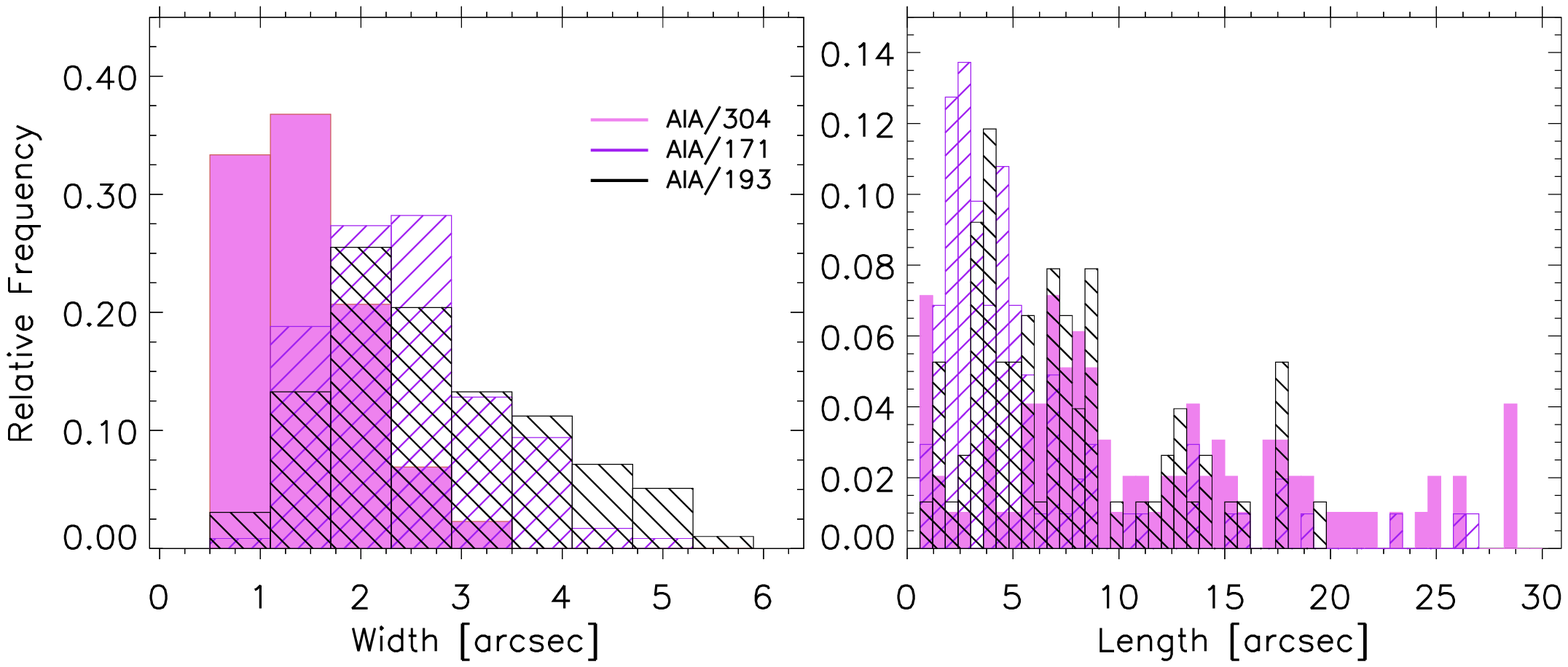}
\end{array}$
\caption{Histograms of calculated widths (\textit{left}) and lengths (\textit{right}) for the detected coronal rain condensations in the various wavelengths used in the present study for datasets 1 and 2. The corresponding colour for each wavelength is shown in the legend. The bin size in each wavelength is set to the spatial resolution of each instrument. For the high resolution instruments (\textit{upper panels}) only the  small scale part of the spatial range is shown. In the \textit{lower panels} only the AIA results are shown. 
\label{fig16}}
\end{center}
\end{figure}

In Fig.~\ref{fig16} histograms displaying the relative frequency (for each wavelength) for the obtained widths and lengths are shown, combining datasets 1 and 2, and therefore all the instruments used in this study (note that the bin widths of the distributions have been set equal to the resolution of each instrument). Due to the large number we restrict the histograms to the more interesting small scales part (\textit{upper panels}) and separate between the high (CRISP, SOT, SJI) and low resolution instruments (AIA in the \textit{lower panels}). The figure clearly shows a significant difference between widths and lengths at all wavelengths. The widths appear nicely distributed, with increasingly higher occurrence frequency at smaller scales for each wavelength and a tail at longer scales. Such distribution shape seems therefore independent of temperature, with an apparent peak for all clumps between $0.2\arcsec$ and $1.1\arcsec$. H$\alpha$ and \ion{Ca}{2}~H clump distributions have, respectively, narrow and broad peaks at $0.25\arcsec$ and $0.6\arcsec$. Blobs observed with \textit{IRIS} in SJI 2796, SJI 1330 and SJI 1400 have similar distributions peaking at $0.6\arcsec-0.8\arcsec$. We note however that SJI 2796 presents a broader distribution for the peak, between $0.6\arcsec$ and $1\arcsec$, probably due to the higher opacity compared to the optically thinner \ion{C}{2} and \ion{Si}{4} lines. For AIA we note that while widths in 171 and 193 peak around $1.8\arcsec - 3\arcsec$, 304 peaks around $0.8\arcsec-1.7\arcsec$.

On the other hand, the lengths show a more random distribution, although with high clustering at small numbers, between $0.5\arcsec$ and $1.5\arcsec$. It is worth noting that the lengths distribution spans up to $30\arcsec$ or so (for instance, the very long clump of Fig.~\ref{fig4}). For lengths longer than $5\arcsec$ (not shown here) the distribution becomes sparse. 

In Table~\ref{table1} average values for each wavelength and dataset are shown, together with the total number of clumps detected in the FOV per minute as an indication for the number of clumps detected in each dataset. We choose to calculate this number since a clump is a not well defined plasma structure over time. As shown in the present movies (and in Fig.~\ref{fig11}) and in Paper~1, clumps can dramatically change shape, appear and disappear along their fall. This number needs to be read with caution: the area over which rain is detected in dataset~1 is roughly half that in dataset~2 (therefore, dividing the number over area would lead to a similar number between \textit{SST} and \textit{Hinode}). The continuous flow in 304 (as opposed to clumpy) also makes it difficult to identify a clump. For AIA only clear cut examples of clumps were selected, which explains the low number per minute. 

The widths presented here in H$\alpha$ are similar to previously measured widths in the same wavelength with CRISP in large statistical datasets \citep{Antolin_etal_2012SoPh..280..457A,Antolin_Rouppe_2012ApJ...745..152A}, but present a relatively higher number of clumps at lower spatial scales. This is mainly due to the fact that the CRISP instrument gained from improved spatial sampling after an upgrade performed in 2009. The widths of \ion{Ca}{2}~H clumps are similar to those reported in \citet{Antolin_2010ApJ...716..154A}. In the top left panel of Fig.~\ref{fig16} the tendency of number increase for clumps with smaller widths suggests a tip-of-the-iceberg scenario. 

As expressed previously, the lengths appear more randomly distributed, especially in H$\alpha$, and therefore may be more dataset dependent. For instance, extremely long clumps are only observed in dataset~1, and have not been previously observed in other datasets in H$\alpha$ with CRISP, except in the recent work by \citet{Scullion_2014ApJ...797...36S} in the context of flares, i.e. stronger footpoint heating sources.

\input{table1}

\section{Discussion}\label{discuss}

\subsection{EUV intensity variations as a signature of catastrophic cooling}

Coronal loops in active regions often present intensity variations, which are regularly associated to states out of hydrostatic equilibrium such as states of thermal non-equilibrium \citep{Aschwanden_2001ApJ...550.1036A, Reale_2010LRSP....7....5R}. A state of thermal non-equilibrium implies the presence of thermal instability in coronal loops. This instability can be thought as the most general state of the plasma (meaning that thermal equilibrium is extremely unlikely). The relevant question here is what are the timescales of this instability in coronal loops? Accordingly, this instability can have a complete character, in the sense that runaway cooling due to the shape of the optically thin loss function may occur locally in coronal loops and continue uninterrupted down to chromospheric temperatures, until heating from local sources overcome the losses due to its own radiation. The completeness of thermal instability is, however, not guaranteed and depends strongly on the efficiency of the heating mechanisms and on thermal conduction.  \cite{Mikic_2013ApJ...773...94M} have shown that even changes in geometry (non-uniformity in area cross-section and asymmetry in the heating) play an important role in determining the aspect of this instability. Furthermore, the spatial scales involved in the cooling processes (for instance, the clumpiness and strand-like structure of the rain) are directly related to the properties of the plasma and to the counter-acting heating mechanisms, and may therefore reveal essential characteristics of the heating processes.

Despite the close link between thermal instability and coronal heating and therefore the importance of thermal instability, we do not know how ubiquitous this instability is in the corona and the details of its character (for instance, whether it is complete or not). This is largely due to the high complexity of the thermal instability in strongly anisotropic plasmas such as the solar corona. Recent observational studies over active regions indicate a general tendency for cooling \citep{Viall_2012ApJ...753...35V}. Although it is clear that not all the loops in an active region are in a thermal non-equilibrium state and undergo catastrophic cooling, it is important to estimate what fraction of loops are.  

The traditional picture of thermal instability implies that coronal loops in a thermal non-equilibrium state become progressively cooler, suggesting a time delay between filters detecting emission at different temperatures. Such cooling progression throughout transition region temperatures has previously been invoked to explain variations in EUV lightcurves \citep{Foukal_1976ApJ...210..575F,Foukal_1978ApJ...223.1046F,Kamio_etal_2011AA...532A..96K,Kjeldseth_Brekke_1998SoPh..182...73K,Landi_etal_2009ApJ...695..221L,Oshea_etal_2007AA...475L..25O,Schrijver_2001SoPh..198..325S,Tripathi_etal_2009ApJ...694.1256T,Ugarte-Urra_etal_2009ApJ...695..642U,UgarteUrra_etal_2006ApJ...643.1245U,Warren_2007PASJ...59S.675W}. However, the direct link to catastrophic cooling has not been yet firmly established, since all of these multi-wavelength observations do not include chromospheric ranges. In this work we make the first steps in this direction by linking commonly found intensity variations in EUV filters of \textit{SDO}/AIA to the observational imprint of catastrophic cooling, i.e. coronal rain observed in chromospheric diagnostics. 

We observe complete thermal instability with temperatures down to the chromospheric range. In Section~\ref{temp} we show that EUV darkening is found to be associated to flows in AIA 304 and presents a continuous and rather persistent character. Such AIA 304 flows are observed along many of the loops within the active region (dataset~1), part of which are captured towards the footpoints in H$\alpha$ by  CRISP. The flows in AIA~304 appear strongly correlated in space to the appearance of cool chromospheric material in the loops observed in H$\alpha$, which occurs in a more intermittent way. This suggests that while the entire loop may be thermally unstable and cooling, plasmas at transition region temperatures are more widespread and common within the loop than plasmas at chromospheric temperatures. This is expected to some extent due to the tendency of the plasma to keep a uniform pressure distribution along the field (and the fact that thermal instability entails a local loss of pressure leading to small condensations).

Small EUV intensity perturbations are further strongly correlated in time with the presence of H$\alpha$ rain clumps. Although intermittent, these clumps can come in series of events and generate quasi-periodic EUV intensity variations on the order of a few minutes. In most cases, these variations are found to be associated to groups of chromospheric clumps which we have previously termed `showers' (Paper~1), and span over a few arcseconds in the transverse direction. Correlation between EUV darkening and H$\alpha$ has been extensively studied in the case of prominences \citep{Heinzel_Anzer_2006ApJ...643L..65H,Labrosse_etal_2010SSRv..151..243L}. For the EUV wavelengths of interest here, this correlation is mainly due to continuum absorption from neutral hydrogen, neutral helium and singly ionised helium. 

\subsection{Multi-thermal character}\label{disc_multi}
 
\begin{figure}
\begin{center}
$\begin{array}{c}
\includegraphics[scale=0.6]{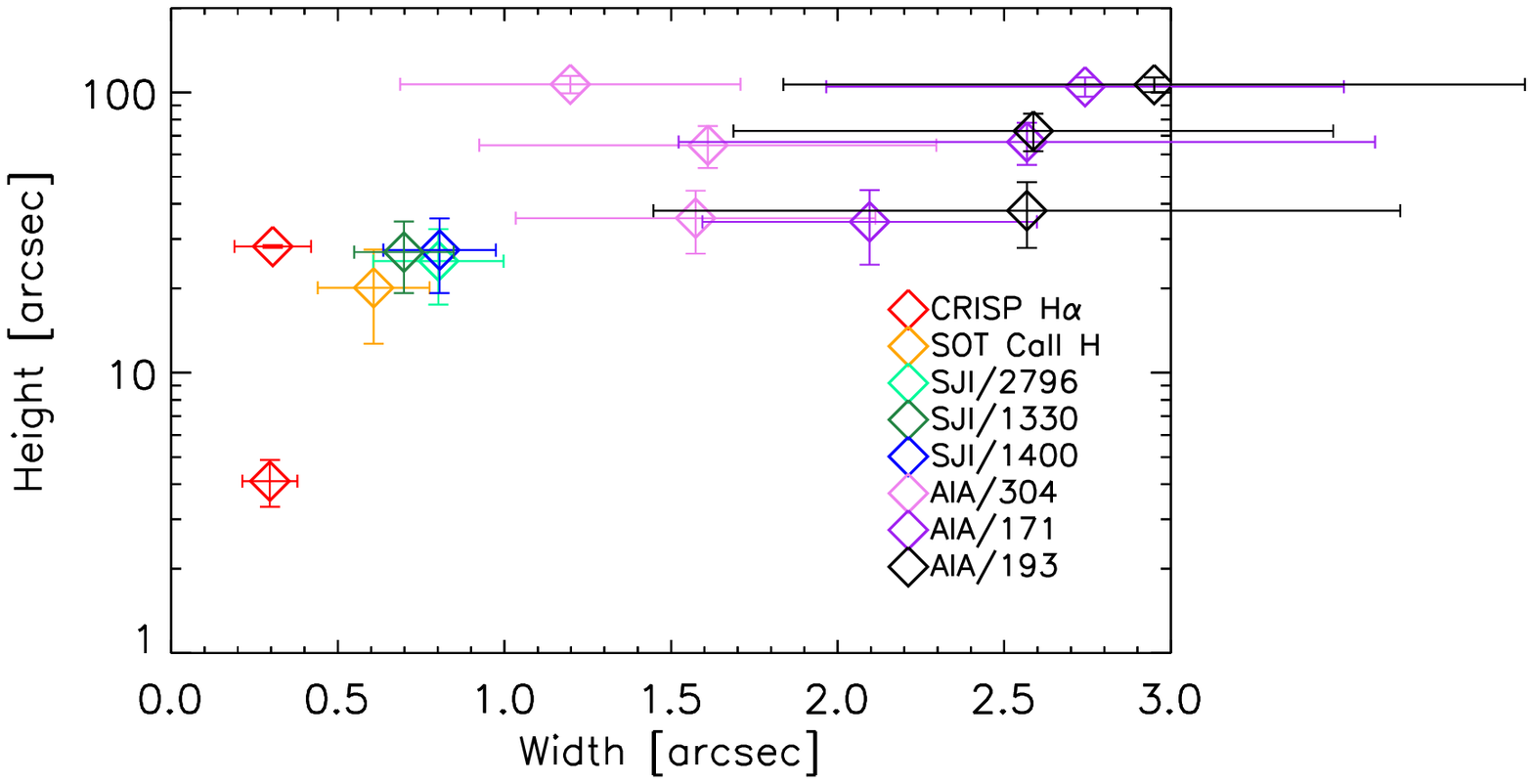} \vspace{-0.15in}\\
\includegraphics[scale=0.6]{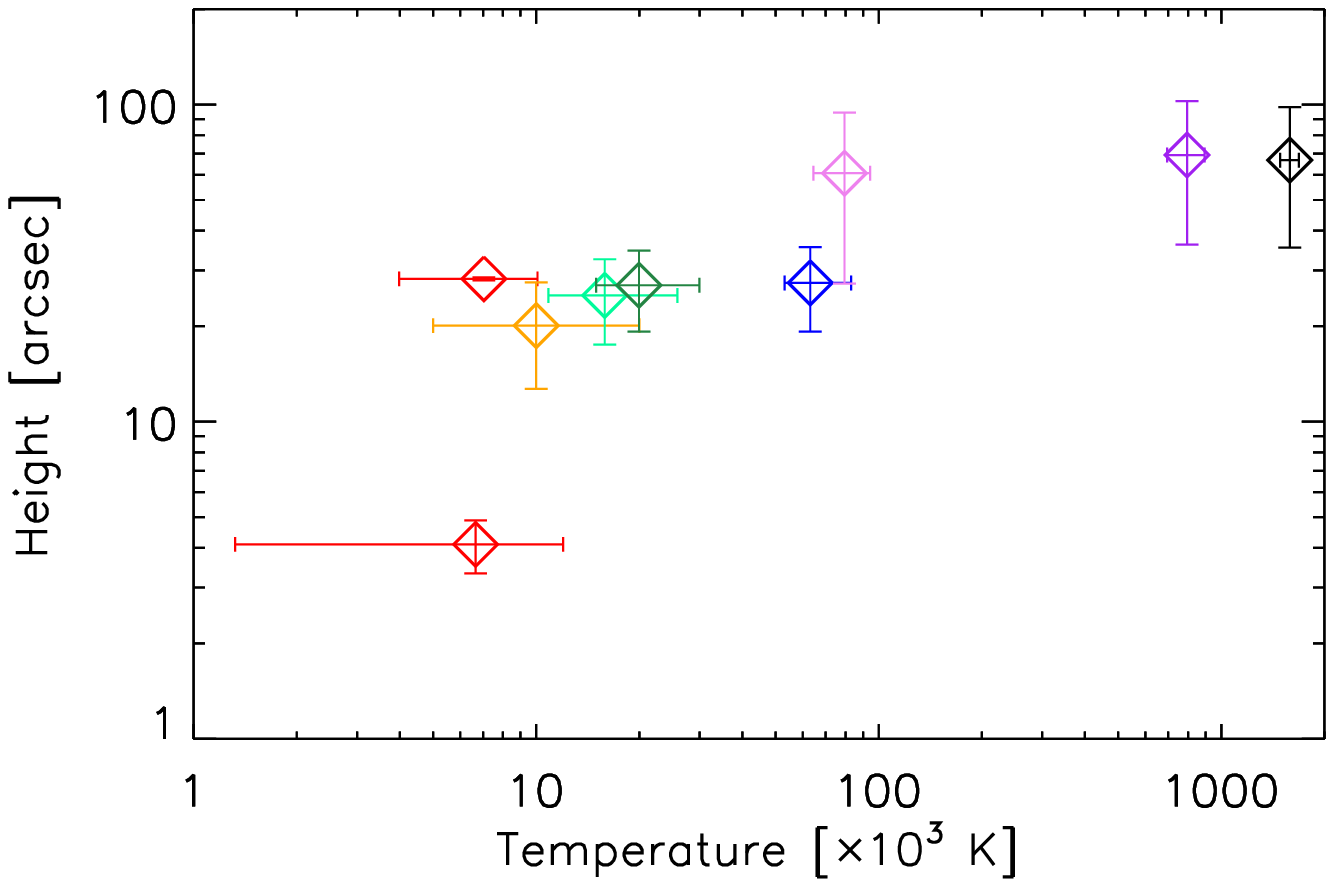} \vspace{-0.15in}\\
\includegraphics[scale=0.6]{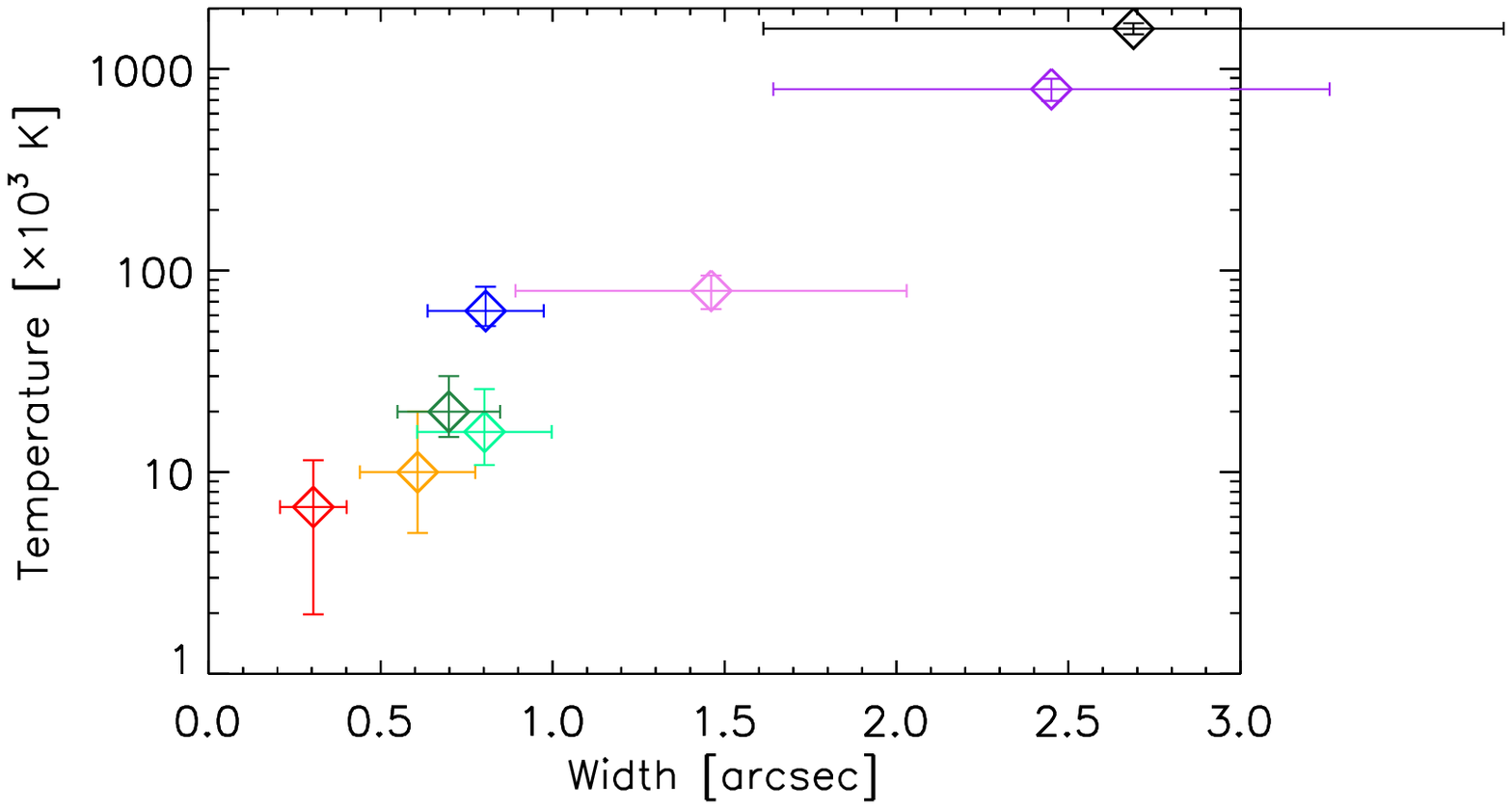}
\end{array}$
\caption{Diagrams showing relationship between various quantities. As quantities we have the height of the measurements, estimated from the footpoint location of the loop in the chromosphere, the width of the condensations and the temperature of the condensations. For each quantity $Q$ its average value $\langle Q\rangle$ over the entire set of measurements is plotted, together with error bars corresponding to the standard deviation $\sigma$ ($\langle Q\rangle\pm\sigma$). Each colour corresponds to a different wavelength, indicated in the top panel.  
\label{fig17}}
\end{center}
\end{figure}

The progressive cooling picture from thermal instability also suggests a difference in emission with height. In this picture, emission in H$\alpha$  and at cooler wavelengths is preferentially detected at low coronal heights. In the middle panel of Fig.~\ref{fig17} we plot the averages of height versus temperature for all detected clumps in datasets 1 and 2. The temperatures taken for the SJI and SOT filters are those of maximum formation for the dominant ions of each filter. Those for CRISP correspond to the measured temperatures, shown in Fig.~\ref{fig6}. Although not strongly correlated, the plot suggests a slight temperature decrease with height. Particularly, the very low temperatures below 5000~K, even down to 2000~K or so, for the H$\alpha$ clumps detected with CRISP at the lower part of the loops provides support to the cooling picture. 

Through co-observation between \textit{IRIS} and \textit{Hinode}/SOT we are able to follow for the first time the rapid progression in the runaway cooling in thermally unstable loops. A two step progression is found in which fast sudden appearance at transition region and upper chromospheric temperatures is followed by a more gradual appearance at lower chromospheric temperatures. This two step scenario agrees with the shape of the optically thin loss function, which predicts runaway cooling down to $10^5$~K or so, below which material becomes progressively optically thick, and therefore radiation has a lower escape probability (leading to less efficient radiative cooling).

Despite the progressive cooling from transition temperatures to chromospheric temperatures found in dataset~2 the AIA 171 intensity along the loop is found to increase about 10 minutes prior to the rain event, and remains bright and even further increases at the end, during the rain event. Such behaviour strongly suggests the presence of multi-thermal plasma in the loop, from chromospheric to coronal temperatures. Furthermore, it implies that the progressive cooling picture down from coronal temperatures to chromospheric temperatures is not always reflected in the evolution of EUV light curves. Hot material can co-exist with catastrophically cooling plasma. Therefore a hot to cool systematic sequence in EUV channels is not an observable requirement for a loop in a thermal non-equilibrium state.

The obtained cooling down to chromospheric temperatures is very fast, achieved in a timescale that varies from tens of minutes (dataset~1) down to minutes (dataset~2). Due partly to this reason, in both analysed datasets, and especially in the one analysed with SJI and SOT, thanks to the high spatial and temporal resolution we find a high degree of co-spatiality in the multi-wavelength emission over the entire span of the event. This strongly supports the presence of multi-thermal plasma in these loops. Furthermore, the multi-thermal character is accompanied by strong density inhomogeneity, which suggests a complex thermodynamic evolution within the same loop. The multi-thermal picture is also supported by the results of \citet{Scullion_2014ApJ...797...36S}, which show H$\alpha$ emission with CRISP co-located with EUV AIA~171 in post-flare loops. 

The large range of temperatures we detect imply that coronal rain, as a result of catastrophic cooling, is not solely a chromospheric phenomenon but a transition region phenomenon as well. Actually, since it is material cooling from coronal temperatures and, as discussed previously, it is more pervasive in transition region temperatures, it is mainly a transition region phenomenon which spans down to chromospheric temperatures if the thermal instability is complete. This result matches well the picture that is being perceived with increasing frequency in SJI and AIA synoptic observations, namely, that basically every active region shows rain-like phenomena in AIA 304 and SJI 1400. A proper statistical analysis of this picture is left as future work.

For dataset~1, the EUV emission (darkening or brightening) in AIA 304 is found to be wider than the chromospheric emission in H$\alpha$ or \ion{Ca}{2}~H suggesting a wide transition from chromospheric to transition region temperatures. However, the highly co-spatial emission obtained from SJI and SOT in dataset~2 suggests that the wide transition in dataset~1 is mainly due to the lack of resolution in AIA. The observed thin transition in dataset~2 must exist on spatial scales below the resolution of \textit{IRIS}, i.e. $0.33\arcsec$, which is also supported by the similar width distributions obtained with SOT and the SJI filters (cf. Fig.~\ref{fig16}). Indeed, when closely comparing the substructure of clumps, small structural differences appear between the clumps in \ion{Ca}{2}~H and in SJI 2796, 1330 and 1400. This multi-thermal plasma picture and the large density inhomogeneities at such high resolution suggests that the thermal instability mechanism in a strongly anisotropic magnetic field such as the corona is far more complex than the simple picture of a uniformly progressive cooling plasma with cool chromospheric cores surrounded by warmer diffuse material. On the other hand, the difference in the average widths between rain in emission in AIA 304 and rain in absorption in AIA 171 and AIA 193 suggests a relatively wide transition from transition region to coronal temperatures on the order of $0.5\arcsec$ \citep[known as Prominence Corona Transition Region, PCTR, in the case of prominences,][]{Chiuderi_1991SoPh..132...81C,Heinzel_2001AA...370..281H}, similar to reported values in prominences \citep{Vial_etal_2012AA...541A.108V,Parenti_2015ASSL..415...61P}. This suggests a large increase in complexity in the interface between chromospheric and transition region plasmas, and less so for the PCTR. It is also possible, although less likely, that since absorption in the AIA~171 filter is produced by multiple elements (but mainly H, He and He$^{+}$), an inhomogeneous He distribution within the loops (more concentrated in the core of the loops than hydrogen) could lead to such difference in widths. How such high degree of inhomogeneity, especially at the smallest scales, is achieved in such loops is an important question that needs to be addressed in future work. Theoretically, the presence of high degree complexity at the smallest scales could be explained by the existence of tangential discontinuities in the field in which material can collapse, predicted by stability analysis of prominence plasma in similar conditions as in the present observations \citep{Low_2012ApJ...755...34L}. Such theoretical studies also predict the constant evolution of magnetic fields and fluid at these small scales leading to spontaneous formation and resistive dissipation of discrete currents \citep{Low_2012ApJ...757...21L}. Such events may explain the constant morphological and physical changes of coronal rain clumps that we observe along their fall, also reported in \citet{Harra_2014ApJ...792...93H}. 

\subsection{Elemental scales and global topology}\label{disc_scales}

The present results strongly suggest that the bulk of the distribution is below the highest resolution used here ($\approx0.2\arcsec$). This study confirms and extends the result of \citet{Scullion_2014ApJ...797...36S} over various instruments spanning different temperatures. This picture is also obtained in 2.5-D MHD simulations of coronal rain by \citet{Fang_2013ApJ...771L..29F}, in which the number of rain clumps keeps increasing at seemingly ever decreasing scales. In Figs.~\ref{fig16} and \ref{fig17} (upper panel) the standard deviation of widths and the ratio between standard deviation and mean values for the widths decreases significantly from the AIA measurements down to CRISP measurements (except for the AIA 304 widths in the second dataset, perhaps reflecting the absence of large scale coronal rain downflows and no significant showers in this case). On the other hand the lengths present a more random distribution with peaks at a few arcsec or less, which is also found in numerical simulations by \citet{Luna_etal_2012ApJ...746...30L} and \citet{Fang_2013ApJ...771L..29F}. We further find extremely long clumps encompassing a few tens of thousands of km. This large scatter in lengths may reflect the presence of several factors at play along the field (such as flows, thermal conduction, instabilities, field  geometry etc.), supported by numerical results. The more uniform distribution of widths, the agreement over a significant number of datasets \citep{Antolin_etal_2012SoPh..280..457A,Antolin_Rouppe_2012ApJ...745..152A,Harra_2014ApJ...792...93H,Scullion_2014ApJ...797...36S} and the similarity with widths of prominence threads \citep{Lin_2005SoPh..226..239L} suggests a more fundamental mechanism behind the shaping of condensations in the transverse direction to the field. This behaviour strongly suggests the existence of elemental strand-like structures at smaller scales. Such fundamental scales would be defined by a similar thermodynamic and magnetic field evolution. A relevant question to ask is then whether thermal instability plays a main role in the shaping of such structures, and especially defining the widths of these elemental structures, or whether the obtained widths would also be expected for substructure in a thermally stable loop. In Fig.~\ref{fig16} spatial resolution and spectral diagnostic (and therefore temperature) are intertwined in the results and it is difficult to disentangle them and estimate the influence of each on the observed widths. The relation between temperature and width is shown in the lower panel of Fig.~\ref{fig17}, where we can see a roughly monotonically increasing trend. However, the fact that H$\alpha$, \ion{Ca}{2}~H and \ion{Mg}{2}~k form at roughly similar chromospheric temperatures but show a significant difference in widths points mostly to a major role of resolution over temperature (although less so from \ion{Mg}{2}~k due to the higher opacity). Also, the widths found in \ion{C}{2} and especially in \ion{Si}{4} are similar, despite the significant difference in temperature formation, suggesting that such small strand-like structures may extend to hotter lines at high spatial resolution. 

It is nonetheless possible that the observed strand-like structure in these loops, or part of it, is the product of thermal instability. For instance, in Section~\ref{morph_str} we showed at the smallest scales we may have an influence of the MHD thermal mode on the shaping of strand-like structure. The ripples in the clump's neighbourhood generated by this wave imply a strand-like structure all around the main clump. These ripples constitute small density enhancements (which correspond to the spatial shape of the thermal mode eigenfunctions) that propagate together with the clump and can enhance radiative losses. It is therefore likely that such perturbations may grow into condensations themselves, leading to further strand generation \citep{Soler_2012AA...540A...7S}. Besides this effect, catastrophic cooling can also introduce large pressure variation, especially in the strong density inhomogeneities found here with cool cores with electron densities on the order of $10^{11}$~cm$^{-3}$. Due to flux freezing the collapse of material following thermal instability generates an increase of the magnetic field. Despite the local character of the condensation the subsequent field enhancement is non-local due to the rigidity of the coronal magnetic field \citep[high magnetic tension due to the high external Alfv\'en speed,][]{An_1984ApJ...276..755A}. This process can thus generate a (slightly) magnetically different strand which would in turn shape warmer material up- and downstream. Such a scenario has been obtained in 2-D MHD simulations of thermal instability in the interstellar medium, with similar plasma $\beta$ values \citep{Sharma_2010ApJ...720..652S}. This scenario implies a relation between clumps' widths and the magnetic field at small scales, which offers interesting coronal seismology applications. 

The proposed non-local enhancement of the magnetic field would increase the magnetic pressure downstream of the clumps, which could explain the less than free-fall speeds (and rather constant accelerations) always observed in coronal rain. Let us make an order of magnitude estimate of this possibility and leave a detailed analysis for future work. Let us consider the case of dataset~1 for which we have density estimates. We have an average velocity at the footpoint of $100~$km s$^{-1}$, a loop height of 42~Mm, and let us assume a negligible initial speed at the apex of the loop. The average effective gravity along an elliptic loop with a ratio of loop height to half baseline between 0.5 and 2 varies between 0.132~km~s$^{-2}$ and 0.216~km~s$^{-2}$ \citep{Antolin_Verwichte_2011ApJ...736..121A}. Taking an average of 0.174~km~s$^{-2}$, the average velocity at the footpoint for our clumps would be around 150~km~s$^{-1}$ due to the effective gravity alone. From Bernouilli's equation we can then estimate the increase of pressure due to the magnetic field downstream from the clump that would lead to a decrease of the final velocity of the clump (assuming no change in the gas pressure downstream of the clump). We obtain values of $0.3-0.5$~dyn~cm$^{-2}$ for average downstream plasma electron densities of $3-5\times10^{9}$~cm$^{-3}$ \citep[a factor of 10 lower than the average density derived from the EUV absorption measurement in Section~\ref{denseblob}, and consistent with numerical simulations,][]{Antolin_2010ApJ...716..154A}. These values correspond to a magnetic field increase of $2.8-3.6$~G. On the other hand, the gas pressure created by the dense clump is $\approx0.38$~dyn~cm$^{-2}$ (assuming a low ionisation degree), matching the obtained values and therefore supporting this possibility.

Besides the interesting possibilities of coronal rain as local probes of the internal thermodynamic and magnetic loop structure, it can also serve as a global tracer of the magnetic field topology. This was first demonstrated in Paper~1, where the angle of fall of rain clumps was calculated based on the full velocity vector, which allowed to trace the lower legs of loops and thus to distinguish several loop families. Here we show that the rain clumps' lengths may exist over a few tens of thousands of megameters, allowing further tracing of the loop legs. Such long rain structures have also been found recently by \citet{Scullion_2014ApJ...797...36S} in the context of post-flare loops, forming from the apex of loops. But even if short, by calculating the variance over relatively long periods of time as in Fig.~\ref{fig14}, or by detecting all the clumps crossing along perpendicular cuts to the trajectory at different heights as in Fig.~\ref{fig5}, the global shape of the loops' magnetic field can be inferred. By doing so we detect a clear expansion in the loops of dataset~1 with CRISP. The width of the loop cross-section over which clumps are detected for loops 1 and 2 in Fig.~\ref{fig5} decreases from $4.6\arcsec$ to $2.4\arcsec$ and $6.8\arcsec$ to $3.4\arcsec$, respectively, from height C down to F separated over a distance of roughly $26\arcsec$ (similar to both loops). The width of the loops at the apex, detected with AIA in datasets 1 and 2 vary around $5\arcsec$, suggesting that the expansions we observe with CRISP are most of those experienced by these loops. On the other hand, in Paper~1 little expansion is observed, perhaps due to the difference in the height at which rain is detected. Here we detect it down to chromospheric heights, just before impact, while in the previous work only detection in the corona was possible. The expansion seems to occur mostly uniformly between heights C and F, with the expansion occurring mostly above an observed kink in the trajectory of the clump, a little above height F. In Fig.~\ref{fig5}, this kink is observed at $y\approx35\arcsec$ (for loop 2 this kink is at a similar height). This kink for loops 1 and 2 is visible whenever a clump is observed and can be followed all the way down, suggesting that it is a feature of the magnetic field topology above the sunspot. Similar kinks in the field above sunspots have been detected by measuring the change with height of the dominant periods of running penumbral waves from the photosphere into the chromosphere \citep{Jess_2013ApJ...779..168J}. 

\subsection{Persistent high-speed downflows above sunspots}\label{disc_pers}

The chromospheric material observed along these thermally unstable loops occurs in clumps and appears intermittently at coronal heights. However, this downflow has a more persistent and constant character at chromospheric heights, just before impact. This is contrary to expected for the following reasons. At low chromospheric heights  projection effects are stronger and visualisation of clumps relies on absorption and therefore on high opacity. Unless the density of clumps is significant the resulting absorption would not be significant enough for allowing detection, which then rules out a significant number of clumps. Also, unless the filters have sufficiently narrow spectral bands the contribution from other chromospheric and photospheric emission will strongly reduce the contrast produced by the clumps. For these reasons, while off-limb detection of clumps can be performed in multiple wavelengths, on-disc detection is strongly limited. Even in the relatively high H$\alpha$ opacity on-disc detection is usually strongly reduced to about 10\% of the rain quantity, as shown in \citet{Antolin_etal_2012SoPh..280..457A} and Paper~1. Here we show that under specific conditions of relatively high Doppler velocities from the H$\alpha$ clumps with respect to the background (and therefore reduced projection effects in the wing of H$\alpha$) detection of clumps just before impact is possible and the occurrence frequency of such downflows increases considerably, enough to give a persistent character.

The increase in coronal rain quantity at low heights may be the combination of two effects. The first effect is the continuous cooling, achieved in the case of complete thermal instability down to chromospheric temperatures. As discussed previously, in this study we obtain evidence for continuous cooling but in a strongly inhomogeneous way in which only small regions of the loops cool down catastrophically. Consequently, more rain in H$\alpha$ will be observed at lower heights. The second effect may be more important, and is strongly dependent on the magnetic field geometry. The increase in coronal rain quantity may be caused by a funnel effect due to the increase of magnetic field strength at lower heights. As discussed in Section~\ref{disc_scales}, here we observe a clear expansion of about a factor of 2 in radius for loops 1 and 2 from heights F to C. On the other hand, only a slight change in the width of the clumps between both heights is found. This is shown in the upper panel of Fig.~\ref{fig17}, where the average of the clumps' widths in H$\alpha$ for both loops changes from $0.305\arcsec$ to $0.296\arcsec$ from cut C to F (shown in Fig.~\ref{fig5}), with nonetheless a significant decrease of the standard deviation from $0.115\arcsec$ to $0.083\arcsec$ respectively. From conservation of mass we would then expect both an increase in the frequency of rain crossing the cuts at chromospheric heights and, correspondingly, an increase of the clumps' lengths with decreasing height. Unfortunately, for the latter, we are unable to provide precise measurements of lengths at lower heights, due to the large projection effects and large speeds of the clumps (as shown in Fig.~\ref{fig14} the height over which the clumps are significantly distinct from the background is limited to the low atmosphere and the region above the spicules). Although this tendency of elongation of clumps has been previously reported \citep{Antolin_2010ApJ...716..154A,Antolin_etal_2012SoPh..280..457A,Antolin_Rouppe_2012ApJ...745..152A} and is observed in general for dataset~2, this quantity is expected to be severely affected by other factors apart from the magnetic field expansion, such as flows, the effective gravity due to curvature, etc. This is reflected in the large scatter in lengths found for all instruments, shown in Fig.~\ref{fig16}. From this funnel effect one would also expect a decrease of the falling speeds from the compression of plasma downstream of the blob, as predicted by numerical simulations \citep{Antolin_2010ApJ...716..154A,Mackay_2001SoPh..198..289M}. In our observations we could not detect a net decrease of falling speeds just before impact. However, it is possible that a velocity decrease is present but cannot be confidently detected due the elongation of clumps and the short distance above the chromosphere over which they offer enough contrast (which leads to a larger uncertainty in the measurements of projected speeds).

Persistent red shifts above sunspots have often been observed in transition region and chromospheric lines \citep{Brekke_etal_1990ApSS.170..135B,Brueckner_1981sars.work..113B,Dere_1982SoPh...77...77D,Kjeldseth-Moe_etal_1988ApJ...334.1066K}, often associated to bright fan-shaped structures observed in EUV above the umbra, usually termed plumes \citep{Foukal_1974ApJ...193L.143F,Brosius_White_2004ApJ...601..546B,Brosius_2005ApJ...622.1216B}. The range of downflow speeds associated to such structures is large, mainly subsonic ($20-40$~km~s$^{-1}$), with supersonic second components of $100~$km~s$^{-1}$ or more \citep{Brekke_etal_1990ApSS.170..135B,Brueckner_1981sars.work..113B,Dere_1982SoPh...77...77D,Gurman_1993ApJ...412..865G,Kjeldseth-Moe_etal_1988ApJ...334.1066K,Nicolas_etal_1982SoPh...81..253N}. Recently, similar events have been detected in transition region lines with \textit{IRIS} above the umbra of a sunspot, reaching speeds of $200~$km~s$^{-1}$ \citep{Kleint_2014ApJ...789L..42K}. Such events, especially those presenting high speed components, have been mainly related to either siphon flows (resulting from pressure imbalance at the footpoints) or condensations flows (issued from thermal instability in the coronal part of the loops). The latter scenario has been more frequently invoked over the years, especially when viewed in chromospheric lines, and since it can also generate siphon flows \citep{Foukal_1978ApJ...223.1046F,Kleczek_1963BAICz..14..167K,Makhumov_etal_1980SoPh...66...89M, Antolin_etal_2012SoPh..280..457A,DeGroof_2004AA...415.1141D,Kawaguchi_1970PASJ...22..405K,Leroy_1972SoPh...25..413L,Muller_2005AA...436.1067M}, also supported by numerical simulations \citep{Goldsmith_1971SoPh...19...86G,Hildner_1974SoPh...35..123H,Mok_etal_1990ApJ...359..228M,Reale_1996AA...316..215R,Reale_1997AA...318..506R,Muller_2003AA...411..605M,Muller_2004AA...424..289M,Muller_2005AA...436.1067M,Mendozabriceno_2005ApJ...624.1080M,Mok_etal_2008ApJ...679L.161M,Antolin_2010ApJ...716..154A,Xia_etal_2011ApJ...737...27X,Fang_2013ApJ...771L..29F}.  

At high resolution, especially in chromospheric lines, coronal rain downflows have been shown to have a clumpy character, thus suggesting a rather intermittent and bursty downflow character, even if frequent, at odds with some observations of persistent and continuous downflows above sunspots. The present observations suggest that the intermittent and clumpy character of cool coronal rain (observed in chromospheric lines) can become persistent and continuous at chromospheric heights, thereby providing an answer to the long standing problem of downflows. When viewed in warmer transition region lines this character may be further enhanced. Indeed, we find that in general the widths of the clumps increase in transition region lines, which may lead to the same effect. In AIA the darkening of the EUV loops is observed to have a persistent and continuous character. The persistency of the rain recently reported in \citet{Kleint_2014ApJ...789L..42K} also supports this scenario.

\subsection{Role in the chromosphere-corona mass cycle}

Here we show that clumps are multi-thermal, suggesting a highly inhomogeneous density structure. Although a thorough statistical study of densities of coronal rain clumps is needed, we can make a rough estimate of the mass density flux involved in the current process for dataset~1. For this we assume that the mass of each condensation is roughly proportional to its length, a result supported by 3D numerical calculations \citep{Luna_etal_2012ApJ...746...30L}. The electron density for a particular clump is estimated to vary between $2\times10^{10}~$cm$^{-3}$ and $2.5\times10^{11}$~cm$^{-3}$. Taking an average of $1.4\times10^{11}$~cm$^{-3}$ for all H$\alpha$ clumps, an average width of $0.4\arcsec$, an average length of $3.8\arcsec$ and taking into account that the total number of H$\alpha$ clumps is 479 for the two loops over a time of 95~min, we obtain an average downward mass flux for one loop of $1.23\times10^{9}$~g~s$^{-1}$. For dataset~2, assuming a similar density and using the corresponding values found for that dataset (cf. Table~\ref{table1}), we find a value of $5.22\times10^{9}$~g~s$^{-1}$. These values are on the same order as that estimated in Paper~1, where, the observed active region comprised multiple rain loops (5 or 6 loops can be distiguished, as shown in their Figs.~14 and 17) and a corresponding higher number of clumps (the assumed average rain density in that paper was however assumed 2.5 times lower than that estimated here). These are significant values comparable to those found for prominence drainage \citep{Liu_etal_2012ApJ...745L..21L,Berger_2012ApJ...758L..37B}.  This agreement over different datasets strongly suggests an important role of coronal rain in the chromosphere-corona mass cycle. 

As discussed in Section~\ref{disc_scales}, the detected number of clumps in H$\alpha$ may very well just be the tip of the iceberg, and many more unresolved clumps may exist. Let's assume nonetheless that we detect most of the rain in both loops over the observed time sequence, which would then correspond to the amount of material in these loops involved in one limit cycle (assuming that most of the material in one loop is thermally unstable). We can then calculate the average density in these loops. For this we assume the loop to be a straight cylinder with an average width of $5.7\arcsec$ (the average between both loops of the transverse distance over which clumps are detected at heights C) and a loop length of 130~Mm, as estimated in Section~\ref{euv2ha} (assuming a similar length for both loops). We obtain an average electron density of $6.2\times10^{8}$~cm$^{-3}$. Using corresponding values for the loop in dataset~2 a value of $7.3\times10^{8}$~cm$^{-3}$ is obtained. These values are at least 5 times below the expected density of active region loops in a thermal non-equilibrium state, obtained from numerical simulations. This suggests three possibilities. In the first and second the loop does not become fully thermally unstable with only parts of it condensing and falling as coronal rain. In the first scenario, most of the loop's mass remains at coronal temperatures and undergoes a different thermodynamical evolution. In the second, the entire loop cools but only a small part of it reaches chromospheric temperatures while most of it stays at transition region temperatures. The third scenario consists of a fully thermally unstable loop, with most of the mass reaching chromospheric temperatures but involving rain clumps smaller than those presently resolved. The first scenario would imply that neighbouring strands do not generally cool down at the same time, something at odds with previous (Paper~1) and the present results (especially for dataset~1, while for dataset~2 it is possible that a significant amount of material at coronal temperatures is still present). In the second scenario, we would expect the bulk of the rain distribution to consist of less dense and warmer clumps being larger and longer (for instance, 3 times larger and longer clumps and twice the currently detected number would suffice). While this roughly matches the AIA statistics for dataset~1, it is however at odds with the \textit{IRIS} results of dataset~2, which show a very similar morphology and distribution of clumps in chromospheric and transition region temperatures, and especially far fewer number of clumps at high temperatures. Hence, the most likely possibility is the third, and involves a far larger number of unresolved chromospheric clumps. This result suggests that the main contributor to downflows (in terms of mass) in thermally unstable loops is coronal rain. This scenario should be tested in 3D numerical simulations.

\section{Conclusions}\label{end}

In this paper two datasets that combine multi-wavelength observations with \textit{SST}/CRISP, \textit{Hinode}/SOT, \textit{IRIS}/SJI and \textit{SDO}/AIA, spanning chromospheric, transition region and coronal temperatures, are used to reveal the multi-strand and multi-thermal aspects of coronal rain at unprecedented detail. We provide important characteristics of the thermally unstable plasma ending up as coronal rain, and more generally of the thermal instability mechanism in the corona. We show how this mechanism can play a significant role into the shaping of coronal structures. Furthermore, we make first steps towards estimating the fraction of coronal plasma in a normal active region undergoing full catastrophic cooling. The results can be listed as follows:

\begin{itemize}[leftmargin=*]
\item While the standard model of progressive cooling is observed, in which light curves peak in order of decreasing temperature, we also find a case in which coronal emission is present and is even enhanced co-temporally and co-spatially with the chromospheric and transition region emission from coronal rain. Loops in a thermal non-equilibrium state may therefore not forcibly show gradual systematic cooling from coronal to chromospheric temperatures, i.e. a sequential EUV intensity variation from hot to cool temperatures (where the hot emission is significantly reduced) is not a necessary observational condition of thermally unstable loops.

\item EUV intensity darkening is found strongly correlated in space to coronal rain appearance. Furthermore, the rain can produce EUV intensity variations on short timescales of a few minutes that can appear quasi-periodic over time.

\item Progressive cooling of coronal rain is observed, leading to a height dependence of the emission. Catastrophic cooling is observed to follow two stages: a rapid cooling throughout transition region temperatures and a slower cooling down to chromospheric temperatures, which may reflect the transition to optically thick plasma states.  

\item Coronal rain is a highly multi-thermal phenomenon encompassing transition region and chromospheric temperatures. We find a high degree of co-spatiality in the multi-wavelength emission implying a strong degree of density inhomogeneity within thermally unstable loops. A thin transition from chromospheric to transition region temperatures must exist on spatial scales lower than $0.33\arcsec$.

\item We estimate coronal rain core electron densities to vary between $2\times10^{10}~$cm$^{-3}$ and $2.5\times10^{11}$~cm$^{-3}$. 

\item The distribution of coronal rain widths is found to be rather independent of temperature with a clear increase of clump number (and a decrease of standard deviation) at small scales for each wavelength. The number of clumps is, however, found to increase sharply across the temperatures towards the chromospheric range, reinforcing the tip-of-the-iceberg scenario proposed in Paper~1. The present results strongly suggest that the bulk of the distribution is below the highest resolution used here ($\approx0.2\arcsec$). 

\item The coronal rain clumps appear organised in strands. Such structure is not limited to chromospheric temperatures but extends at least to transition temperatures as well, suggesting an important role of thermal instability in the shaping of fundamental loop substructure. At the smallest detected scales we find structure reminiscent of the MHD thermal mode \citep{Field_1965ApJ...142..531F,VanderLinden_1991SoPh..134..247V,VanderLinden_1991SoPh..131...79V, Murawski_etal_2011AA...533A..18M}. 

\item Through order of magnitude estimates we show that the strong density inhomogeneity that we find entails a local increase of gas pressure accompanied by a non-local increase of magnetic pressure due to flux freezing all along the clump's trajectory (up- and downstream), thereby shaping hotter material into strands. We further suggest that this mechanism may explain the commonly observed lower than free-fall speeds of coronal rain.

\item We show how coronal rain can serve as a global tracer of the magnetic field topology. We detect the presence of a clear kink in the field in the upper chromosphere above sunspots, reminiscent of the sharp change in angle reported in \citet{Jess_2013ApJ...779..168J} in a similar location. The magnetic field is found to expand roughly uniformly above this kink up to coronal heights over at least a factor of 2 in radius.

\item The intermittent and clumpy appearance of coronal rain at coronal heights becomes more continuous and persistent at chromospheric heights just before impact. This is likely due to progressive cooling of the rain and, especially, to a funnel effect from the expansion of the magnetic field at low heights. 

\item The persistent (and burst-less) character of coronal rain at low heights is in agreement with the common detection of moderate and high speed red shifts above sunspots in chromospheric and transition region lines, and may thus offer an explanation to this long standing problem.

\item In terms of filling factor (and therefore detectability, especially with poor resolution instruments), coronal rain is mainly a transition region phenomenon. However, in terms of mass, the chromospheric component of the cooling plasma plays a major role in the mass cycle of the loop. We find average downward mass fluxes per loop of $1-5 \times10^{9}$~g~s$^{-1}$, in accordance with previous estimates in Paper~1, and prominence mass drainage values \citep{Liu_etal_2012ApJ...745L..21L,Berger_2012ApJ...758L..37B}. 

\end{itemize}

\acknowledgments
P.~A. would like to thank W. Liu, B.C. Low, B. De Pontieu, R. Soler and the anonymous referee for fruitful comments and discussions that lead to a significant improvement of the manuscript. P.~A. acknowledges support from the International Space Science Institute, Bern, Switzerland to the International Team on `Implications for coronal heating and magnetic fields from coronal rain observations and modeling'. This research was supported by Japan Society for the Promotion of Science (JSPS) of Grant-in-Aid for Scientific Research (A) (Grant Number 25220703, PI: S. Tsuneta). G.~V., T.~P. and L.~R. acknowledge funding by the Norwegian Research Council, and G.~V., T.~P. also by the European Research Council under the European Union's Seventh Framework Programme (FP7/2007-2013)\,/\,ERC grant agreement nr.~291058. E.~S. is a Government of Ireland Post-doctoral Research Fellow supported by the Irish Research Council (IRC). \textit{IRIS} is a NASA small explorer mission developed and operated by LMSAL with mission operations executed at NASA Ames Research Center and major contributions to downlink communications funded by the Norwegian Space Center through an ESA PRODEX contract. The \textit{SST} is operated on the island of La Palma by the Institute for Solar Physics of Stockholm University in the Spanish Observatorio del Roque de los Muchachos of the Instituto de Astrof{\'\i}sica de Canarias. \textit{Hinode} is a Japanese mission developed and launched by ISAS/JAXA, with NAOJ as domestic partner and NASA and STFC (UK) as international partners. It is operated by these agencies in cooperation with ESA and NSC (Norway). \textit{SDO} is part of NASA's Living With a Star Program. This work was (partly) carried out on the Solar Data Analysis System operated by the Astronomy Data Center in cooperation with the Hinode Science Center of the National Astronomical Observatory of Japan.

\bibliographystyle{aa}
\bibliography{ms.bbl}  

\end{document}